%% file: main.tex
\documentclass[%
 reprint,
 amsmath,amssymb,
 aps,
]{revtex4-2}

\usepackage{graphicx}
\usepackage{dcolumn}
\usepackage{bm}
\usepackage{graphicx}
\usepackage{subfigure}
\usepackage{xcolor}         
\usepackage{comment}
\usepackage{hyperref}
\usepackage{layouts}
\usepackage[printonlyused]{acronym}
\input{acronyms}
\usepackage[capitalise]{cleveref}
\usepackage{comment}

\newcommand{\changed}[1]{\textcolor{black}{#1}}

\makeatletter
\AtBeginDocument{
   \def\ltx@label#1{\cref@label{#1}}
   \def\label@in@display@noarg#1{\cref@old@label@in@display{#1}}
\def\label@in@mmeasure@noarg#1{%
    \begingroup%
      \measuring@false%
      \cref@old@label@in@display{#1}
    \endgroup}%
 } %
\makeatother

\begin{document}

\preprint{APS/123-QED}

\title{ Joint inference for gravitational wave signals and glitches using a data-informed glitch model}

\author{Ann-Kristin Malz}
\affiliation{%
 School of Physics and Astronomy,
 University of Glasgow,
 Glasgow G12 8QQ, United Kingdom
}
\affiliation{%
 Centre for Particle Physics \& Astronomy,
 Royal Holloway University of London,
 Egham TW20 0EQ, United Kingdom
}
\author{John Veitch}%
\affiliation{%
 School of Physics and Astronomy,
 University of Glasgow,
 Glasgow G12 8QQ, United Kingdom
}

\date{\today}

\begin{abstract}
    Gravitational wave data are often contaminated by non-Gaussian noise transients, ``glitches'', which can bias the inference of astrophysical signal parameters. Traditional approaches either subtract glitches in a pre-processing step, or a glitch model can be included from an agnostic wavelet basis (e.g. BayesWave). In this work, we introduce a machine-learning-based approach to build a parameterised model of glitches. We train a normalising flow on known glitches from the Gravity Spy catalogue, constructing an informative prior on the glitch model. By incorporating this model into the Bayesian inference analysis with Bilby, we estimate glitch and signal parameters simultaneously. We demonstrate the performance of our method through bias reduction, glitch identification and Bayesian model selection on real glitches. Our results show that this approach effectively removes glitches from the data, significantly improving source parameter estimation and reducing bias. 

\end{abstract}

\maketitle

\section{\label{sec:intro}Introduction}

\input{1.introduction}

\subsection{Related work}
\input{2.related_work}

\section{Background}\label{sec:background}
\input{2.background}

\section{Method}\label{sec:method}
\input{3.method}

\section{Analysis}\label{sec:analysis}
\input{4.analysis}

\subsection{Glitch-only fitting}\label{sec:glitch-only}
\input{5.glitch-only_fitting}

\subsection{Signal-only}\label{sec:signal-only}
\input{6.signal-only}

\subsection{Injection tests}\label{sec:injections}
 \input{7.injection-tests}

\subsection{Bias tests}\label{sec:bias}
\input{8.bias-tests}

\section{Discussion and Conclusions}\label{sec:conclusion}
\input{10.summary}

\section*{Acknowledgments}
We are grateful for feedback on this work from members of the LVK Collaboration; and in particular Christopher Berry, Laura Nuttall, Cailin Plunkett and Rhiannon Udall. We are also grateful for useful discussions and technical advice from Michael Williams and Daniel Williams. JV was supported by STFC grant ST/V005634/1. We are grateful for the computational resources provided by Cardiff University, and funded by STFC awards supporting UK Involvement in the Operation of Advanced LIGO.

This research has made use of data or software obtained from the Gravitational Wave Open Science Center (gwosc.org), a service of the LIGO Scientific Collaboration, the Virgo Collaboration, and KAGRA. This material is based upon work supported by NSF's LIGO Laboratory which is a major facility fully funded by the National Science Foundation, as well as the Science and Technology Facilities Council (STFC) of the United Kingdom, the Max-Planck-Society (MPS), and the State of Niedersachsen/Germany for support of the construction of Advanced LIGO and construction and operation of the GEO600 detector. Additional support for Advanced LIGO was provided by the Australian Research Council. Virgo is funded, through the European Gravitational Observatory (EGO), by the French Centre National de Recherche Scientifique (CNRS), the Italian Istituto Nazionale di Fisica Nucleare (INFN) and the Dutch Nikhef, with contributions by institutions from Belgium, Germany, Greece, Hungary, Ireland, Japan, Monaco, Poland, Portugal, Spain. KAGRA is supported by Ministry of Education, Culture, Sports, Science and Technology (MEXT), Japan Society for the Promotion of Science (JSPS) in Japan; National Research Foundation (NRF) and Ministry of Science and ICT (MSIT) in Korea; Academia Sinica (AS) and National Science and Technology Council (NSTC) in Taiwan.

\bibliography{bibliography}

\end{document}

%% file: acronyms.tex
\newacro{GW}{Gravitational Wave}
\newacro{NF}{Normalising Flow}
\newacro{SVD}{Singular Value Decomposition}
\newacro{LVK}{LIGO-Virgo-KAGRA}
\newacro{ML}{Machine Learning}

%% file: 1.introduction.tex
Since the first detection~\cite{abbott2016GW150914}, the field of \ac{GW} astronomy has now progressed toward regular observations. During the first three observing runs, around 100 signals were detected~\cite{GWTC3} by the interferometers of the \ac{LVK} collaboration~\cite{aasi2015advancedligo, acernese2014advancedvirgo, aso2013kagra}. The sensitivity of the detectors is limited by background noise, comparable in amplitude to a \ac{GW} signal from an astrophysical source, approximated as quasi-stationary coloured Gaussian noise as well as non-Gaussian noise transients known as \textit{glitches}~\cite{Abbott2020ligovirgonoise}. Much of the background noise can be filtered from the data, for example, by excluding frequencies of known noise sources such as the electrical power grid~\cite{Abbott2020ligovirgonoise} and filtering frequencies outside the expected range of astrophysical signals. 

Glitches, however, are not filtered from the data by these generic methods and require special attention. They are troublesome as they can be mistaken for astrophysical signals or bias the parameter estimation results of detected signals. Additionally, they often have unknown physical origins and/or are difficult to mitigate in the detectors~\cite{davis2022detectorCharacterisationMitigation}, while occurring at a rate of approximately one per minute~\cite{abbott2023gwtc3}. 

Glitches are found in the data by Omicron~\cite{robinet2016omicron, Robinet:2020lbf}, which searches the GW strain data of each detector for excess power and characterises properties such as frequency, duration, and amplitude of each glitch. There are a variety of different glitch classes, with different underlying causes and distinct features in the data. The classification of glitches depending on features in the time-frequency domain is addressed by the Gravity Spy project for LIGO~\cite{zevin2017gravityspy, Zevin:2023rmt} and GWitchHunters for Virgo~\cite{razzano2023gwitchhunters}.

In the third observing run of LIGO and Virgo, around $20\%$ of the \ac{GW} signals detected contained glitches that had to be mitigated~\cite{Davis:2022ird}. To improve the detection of \ac{GW}s, the causes of the glitches must be identified and eliminated, or, alternatively, the glitches removed from the data. \changed{It has been shown that bias introduced to signal parameter estimation due to glitches can be mitigated by removing the glitches from the data pre analysis~\cite{Ghonge:2023ksb}}. 

Glitch mitigation is an area of increasing interest as the number of detected gravitational wave signals grows. The third observing run detected several signals~\cite{GWTC2,GWTC2.1,GWTC3} where glitches had to be removed before source parameter inference~\cite{Davis:2022ird}. The earliest and clearest example of a glitch impacting analysis was GW170817, the first binary neutron star detection~\cite{abbott2017gw170817}. In this case, the glitch was sufficiently simple that it could be subtracted by a manual procedure from the data, followed by regular inference of the source parameters~\cite{LIGOScientific:2018hze}. This iterative subtraction procedure works well if there is little interference between the signal and the glitch (i.e. a low inner product between the glitch and the signal waveform), and the glitch is easily modelled, but in the general case one would rather perform simultaneous fitting of signal and glitch, accompanied by statistical uncertainty on the parameters of both. Examples of signals where the glitch removal was less clear-cut are GW191109~\cite{GWTC3, Davis:2022ird, Udall:2024ovp} and GW200129~\cite{GWTC3, Davis:2022ird, Payne:2022spz, Macas:2023wiw}.

When using a traditional stochastic sampler (MCMC or nested sampling) for performing inference, one requires a parametric model of the glitches that can be included as part of the model of the data, possibly in addition to a signal. Such a model could be based on a very agnostic, phenomenological description, such as wavelets, or on a physical understanding of the glitch generation process (e.g. scattered light). From a Bayesian point of view, these correspond to different strengths of prior information assumed in the model. 

In this work, we develop a model that lies between these extremes. By examining known glitches from the Gravity Spy dataset~\cite{Glanzer:2022avx} we can find their dominant components, and train a \ac{NF} to describe the joint prior distribution of these components. In this way, our method incorporates physical information from previous glitches while retaining flexibility in its parameterisation. We demonstrate that this model can successfully mitigate parameter estimation bias caused by glitches.

%% file: 2.related_work.tex
There has been much work on analysing and mitigating glitches in \ac{GW} data, using a variety of tools and methods. Some of these focus on modelling only the glitch in order to detect and flag these times, while others aim to include glitch models as part of general inference. Below, we review similar works in the field and their relationship to our own method.

BayesWave~\cite{cornish2015bayeswave} is the most established glitch removal algorithm in use in the \ac{LVK} collaboration and is routinely used when glitches contaminate data surrounding a detected signal. BayesWave is a Bayesian inference-based algorithm where non-Gaussian features (glitches and signals) are modelled as a sum of Morlet-Gabor (sine-Gaussian) wavelets. The total model can contain a variable number of wavelet components, whose parameters are sampled using trans-dimensional jump Markov Chain Monte Carlo, giving it great flexibility at identifying generic transients. BayesWave has thus proven very useful at modelling and removing glitches from the strain data, however, there are limitations. One disadvantage is that the glitch model constructed from the wavelets does not ``know'' anything about real glitches, but simply models the patterns observed in the data.~\citet{Davis:2022ird} discusses this further, and highlights the importance and need for different glitch mitigation methods suitable for the wide range of possible glitch and signal scenarios.
\citet{Plunkett:2022zmx} simultaneously estimate compact binary and noise parameters using the BayesLine algorithm for Power Spectral Density (PSD) estimation~\cite{Littenberg:2014oda}, as well as the BayesWave sine-gaussian wavelet glitch model. Simultaneous glitch and signal modelling with BayesWave is described in Refs.~\cite{Chatziioannou:2021ezd, Hourihane:2022doe}. 

The other primary glitch removal method used in the third observing run is gwsubtract~\cite{davis2019improvingsensitivity}. The algorithm uses information from an auxiliary witness channel (correlated with, but independent of the strain data) to model and subtract glitches. 

\citet{Ashton:2022ztk} use a Gaussian Process to model long-duration glitches, and perform inference on a combined signal + glitch model by including the Gaussian Process in the Bayesian noise likelihood, although the glitch model is not informed by data.
\citet{Merritt:2021xwh} developed a glitch model (`Glitschen') based on probabilistic principal component analysis for dimensional reduction, followed by a multivariate normal model for the resulting components. This allows the glitch prior to be informed by the set of observed glitches, similar to our approach; however, the prior is limited to those that can be captured by the multivariate normal distribution. Unlike our work, it does not integrate a signal model in the analysis.
\citet{Mohanty:2023mjn} make use of the adaptive spline fitting method SHAPES to subtract broadband, short-duration glitches (Blip, Koi Fish, and Tomte~\cite{zevin2017gravityspy}). They inject a binary neutron star signal overlapping the glitch to imitate a GW170817-like case, and also apply their model to GW170817 directly. 
\citet{Udall:2024ovp} implement a physically motivated model for scattered light glitches and perform joint Bayesian inference of the glitch and signal, extending the work done in~\cite{Udall:2022vkv}. 

The advent of \ac{ML} methods has led to a range of novel approaches to glitch identification and mitigation. Gravity Spy~\cite{zevin2017gravityspy, Zevin:2023rmt} is a citizen-science project that leverages \ac{ML} to classify glitches by their morphological features. The algorithm consists of a convolutional neural network~\cite{Bahaadini:2018git, Soni:2021cjy, wu2024advancingGravitySpy}, which is trained on time-frequency-energy plots, known as omega scans or Q-transforms~\cite{chatterji2004multiresolution}, that have been manually classified by experts and volunteers. Both citizen scientists and the trained \ac{ML} algorithm contribute to the classification of new glitches. In our work, we use the classifications made by Gravity Spy to build a training dataset of glitches.

\citet{Vajente:2019ycy} developed a \ac{ML} method to model noise transients in the strain channel based on information from auxiliary channels, and demonstrated a reduction in parameter estimation bias when glitches were subtracted before inference.
\citet{Wang:2022quo} develop a deep neural network approach, WaveFormer, for noise and glitch suppression in \ac{GW} data, resulting in a cleaned data stream for further analyses.
\citet{Macas:2023wiw} use a neural network to subtract broadband noise in the Livingston detector observed around the signal GW200129. Their network was trained to model the strain data $h(t)$ as a function of three auxiliary channels.
\citet{Lopez:2022lkd} use Generative Adversarial Networks (GAN), to learn the distribution of blip glitches and to generate artificial populations. Similarly, \citet{Dooney:2024pvt} use a GAN, cDVGAN, to create a model that can be used to generate Blip and Tomte glitches and binary black hole signals.
\citet{Li:2024wlt} create an unsupervised model using an Autoencoder, CTSAE, to cluster glitches and identify glitch classes similarly to Gravity Spy.
\citet{Bondarescu:2023jcx} present a quasi-physical, four-parameter, glitch model, Antiglitch, to model and thus subtract Blip, low-frequency Blip, Tomte and Koi Fish glitches.
\citet{Sun:2023vlq} use a normalising flow for likelihood-free inference of GW source parameters, based on training data that include Blip and Scattered Light glitches. In \citet{Xiong:2024gpx}, the authors then extend the work to not rely on glitch modelling by training on signals in Gaussian noise. Applying it to signals contaminated by injected glitches modelled as sine-Gaussians, they show that the posterior from their normalising flow is less biased compared to the standard inference code {\tt Bilby}~\cite{Ashton2019Bilby, Romero-Shaw:2020owr}. 
\citet{Legin:2024gid} use a score-based model instead of the traditional Whittle likelihood to relax the assumption of Gaussian noise in doing parameter estimation.

Our approach has a similar motivation to some of the above works - the desire to remove glitches from the data and thus improve inference - but we offer some key differences. Primarily, we implement our glitch model into the existing, well-established signal analysis code {\tt Bilby}. Except for \citet{Ashton:2022ztk} and \citet{Udall:2024ovp}, the above-mentioned papers develop independent workflows - instead building on the established code simplifies the usage of the method for downstream users. Furthermore, by analysing glitch and signal simultaneously with {\tt Bilby}, our work can be easily integrated into \ac{LVK} analysis workflows. Secondly, by training on real glitches, our model can capture the variety of features present within a glitch class, and the result of the training is a proper Bayesian prior on the glitch parameters. Thirdly, while we mainly focus on demonstrating the robustness of our model on Blip glitches in this paper, the method is applicable to any type of glitch, provided a new model is trained.

This paper is structured as follows; \cref{sec:background} briefly introduces \ac{NF}s and Bayesian inference, \cref{sec:method} details our glitch-model-and-subtracting method, and in \cref{sec:analysis} we perform various analysis to test our model.~\cref{sec:conclusion} discusses our conclusions.

%% file: 2.background.tex
\subsection{Normalising Flows}\label{sec:NF}
Normalising Flows (\acp{NF})~\cite{NFIntroandRevofCurrentMethods,jimenez2015variational, papamakarios2021normalizing} are a type of generative \ac{ML} algorithm, which can explicitly learn the probability density function of given data. A \ac{NF} maps a complex distribution $p_x(x)$, such as a set of data, to a simple latent distribution $p_z(z)$, such as a uniform or Gaussian distribution, by applying one or multiple transformation functions. These functions $f$ are bijections, and the inverse can thus be used to generate a realisation of the data, such that $z=f(x)$ and $x=f^{-1}(z)$. 
Since the transformation is built by composing many layers of simple transformations, it is possible to evaluate its Jacobian determinant,
so that the probability distribution of the data can be defined as, 
\begin{equation}
    p_x(x) = p_z(f(x)) \left |\texttt{det}\left ( \frac{\partial f(x)}{\partial x} \right ) \right |,
    \label{eq:change_of_var}
\end{equation}
where $\partial f(x) / \partial x$ is the Jacobian of the transformation function $f$.
With many layers of transformation, a complex distribution can be approximated by a simple one. The \ac{NF} is trained to learn the transformation function $f$ and the Jacobian determinant, which can be done by minimising the negative log-likelihood (Kullback-Leibler divergence) of the data.
Applying the inverse transform $f^{-1}$ of the trained flow to the latent space $p_z$, samples of the data distribution $p_x$ can be obtained. 
There are different types of \ac{NF}s, with different properties, as described in \cite{NFIntroandRevofCurrentMethods}. Autoregressive flows are generally more expressive but computationally expensive to train, while coupling flows are less flexible but computationally efficient. In this work, we chose the latter for their efficiency.

\subsection{Bayesian inference}\label{sec:bayesian_inference}
In this work, we adopt a Bayesian approach to data analysis. Bayesian inference describes how prior knowledge can be updated when observations provide new evidence through Bayes theorem,
\begin{equation}
    p(\theta|d,H) = \frac{p(d|\theta,H)p(\theta|H)}{P(d|H)},
    \label{eq:bayes}
\end{equation}
which defines the probability density as a function of parameters $\theta$, given the observed data $d$ and postulated model $H$. Here the posterior $P(\theta|d,H)$ depends on the likelihood $P(d|\theta,H)$, the prior $P(\theta|H)$, and the evidence $P(d|H)\equiv\mathcal{Z}$.

The likelihood describes the probability of the data for the model given the parameters and is chosen based on the problem, as it depends on the noise model. For \ac{GW} analysis, Gaussian noise is usually assumed~\cite{thrane2020BayesianInferenceIntro}. In this project, we build a glitch model and implement it as the Bayesian prior distribution over the glitches. The likelihood can then be calculated from the data when modelled as including a glitch, and optionally a signal. If a signal $s(t)$ is present, the data $d(t)$ can be expressed as $d(t)=s(t)+n(t)+g(t)$, where $n(t)$ is the Gaussian noise and $g(t)$ the glitch. 

Besides parameter estimation, Bayesian inference can also be used for model selection. The Bayesian evidence $\mathcal{Z}$ is the normalisation factor in Bayes theorem, and can be written as 
\begin{equation}
   \mathcal{Z} = \int p(d|\theta,H)p(\theta|H) \,d\theta.
    \label{eq:bayesian_evidence}
\end{equation}
Comparing the evidence for different models indicates which model is statistically preferred. The Bayes factor is the evidence ratio of two different models, and can thus be used to compare, for example, a signal model to a Gaussian noise model to determine if there is a signal present. The log of the Bayes factor, $\log \text{BF} = \log(\mathcal{Z}_1) - \log(\mathcal{Z}_2)$, is used for convenience in this paper.

%% file: 3.method.tex
The aim of this work is to model and remove glitches from the LIGO \ac{GW} data simultaneously with the signal analysis being performed. We accomplish this by training a \ac{NF} on glitches in the Gravity Spy O1 training dataset~\cite{coughlin_2018_1486046}, and implementing the trained \ac{NF} model into the Bayesian \ac{GW} inference library {\tt Bilby} \cite{Ashton2019Bilby} as a prior for the glitches. Thus, we can perform a single analysis on a \ac{GW} event, where the glitch is removed and the signal is analysed. 

The work consists of two parts, training and analysis. During training, we train a \ac{NF} to build the glitch model. The \ac{NF} can be trained to learn the shape of the glitches and the transformation between the glitch model and a Gaussian distribution. The Gravity Spy project provides data such as the time and duration of individual glitches for many different glitch classes. For this initial study, we primarily focus on Blip glitches, characterised by their short time duration ($\sim10$ms) and large frequency bandwidth ($\sim100$Hz). A typical Blip glitch is shown in \changed{\cref{fig:glitchonly} (time-domain) and} \cref{fig:glitchonly_qscan} \changed{(time-frequency spectra)}. Blip glitches occur in both LIGO detectors, with an average of two Blips per hour, and are of high interest due to their shape resembling the signal of high-mass compact binary mergers and due to their origins being largely unknown~\cite{cabero2019blipglitches}. The Blip glitches were chosen as the first glitch type to model, due to their short duration and characteristic shape, simplifying the analysis. Our method is straightforward to apply to other glitch classes, and a brief investigation shows similar performance (see~\cref{sec:conclusion}). However, we will leave a full investigation of other glitch classes to future work. 

\subsection{The Training Data}
The \ac{GW} strain data are publicly available from the Gravitational Wave Open Science Centre (GWOSC)~\cite{gwosc} and were accessed using the \texttt{GWpy} python package~\cite{gwpy}. Using GWOSC data, we create a training dataset consisting of 1 second of strain data around each of the Blip glitches in the Gravity Spy training set with durations $<1$ second (a total of 1785 glitches). The data are filtered to remove noise from outlying frequencies, using a bandpass filter, and whitened to normalise the power at all frequencies so that the glitches become clearer. Next, we cut the strain data to the desired length, and then we apply a Hann window to ensure the model tapers smoothly to zero at the limits of its time stretch. See \cref{sec:choosing_filtering} for further discussion.

Having compiled our training dataset, we use \ac{SVD} to reduce its dimensionality, making it simpler to describe with the \ac{NF}.
\ac{SVD} factorises the data set into three separate matrices such that $A_{m\times n}=U_{m\times m}S_{m\times n}V^T_{n\times n}$, where $U$ and $V^T$ contain the orthonormal eigenvectors of $AA^T$ and $A^TA$ respectively, and $S$ is a diagonal matrix consisting of the singular values (the square roots of the eigenvalues of $A^TA$) along the main diagonal. 
Combining $U$ and $S$ such that $T=US$, the training data set of glitches, $G$, can be described by $G=TV^T$, where $V^T$ is the basis matrix containing the time series of all the glitch components, and $T$ represents the weights, or glitch amplitudes, assigned to each basis. Thus, the \ac{NF} can be trained on $T$ only, and the glitch is later reassembled using the saved $V^T$. 
The dimensions of $T$ can be reduced by only keeping the most significant eigenvectors. The cut-off can be obtained by limiting the power loss allowed.
Here the cut-off was chosen so that 97\% of the total power is contained in the remaining bases so that each glitch can be described by around 10 parameters, depending on the filtering and frequency bands used.

\subsection{Choosing the model}\label{sec:choosing_filtering}
Glitches from the Gravity Spy training set~\cite{coughlin_2018_1486046} from the first observing run, O1, were used as the training data for this paper. Glitches identified by Gravity Spy in O1, O2, O3a, and O3b~\cite{Glanzer:2022avx, glanzer_2021_5649212} were used to test the model. As the model is easy to retrain, other training datasets could also be explored. 

The data were filtered to remove noise from outlying frequencies, using a bandpass filter between 20 and 400 Hz. The bandpass was chosen as the majority (92\%) of the glitches in the training set have peak frequencies within this range, and we wish to exclude higher frequencies where lines in the noise spectrum due to e.g. violin modes produce additional features that are imperfectly removed by whitening with a representative spectrum (see below). With further data treatment, the frequency range may be adapted in the future to include a different range of frequencies depending on the aim of the analysis and the properties of the targeted glitches. However, the number of bases needed to model a glitch increases with a broader bandwidth, and an increased number of parameters in the glitch model entails longer run times when applying the model to data. We find that this range is suitable for the glitches considered in this initial study.

The data are then whitened before training the normalising flow, using the O1 representative amplitude spectral density (ASD) for each of the LIGO detectors~\cite{gwosc}. This normalises the power at all frequencies, reducing the noise level and scaling the data to values of order one. The latter is important to improve the performance of the machine learning algorithm. Since the glitch model is thus whitened, we need to un-whiten it before applying it to a glitch. This is done using the ASD calculated from the real data used in the inference analysis, so that the glitch model matches the data properties.

Another consideration is the length in time of the model, as a longer model also leads to more bases. For this project, Blip glitches with durations up to 1 second (according to the Omicron~\cite{robinet2016omicron, Robinet:2020lbf} values reported in the Gravity Spy data set~\cite{coughlin_2018_1486046}) were chosen. However, the actual peaks of the glitches are significantly shorter than this duration, generally around $10$ ms long. Thus, the data chunks can be shortened further to 1/8\,s, reducing the number of model parameters and decreasing run times. After applying the Hann window to these reduced-length data chunks, this still leaves the model long enough to effectively capture the Blip glitches considered here. 
Choosing the time and frequency constraints as discussed, we obtain a glitch model that is described by 12 parameters. 

\begin{figure*}
    \includegraphics[width=\textwidth]{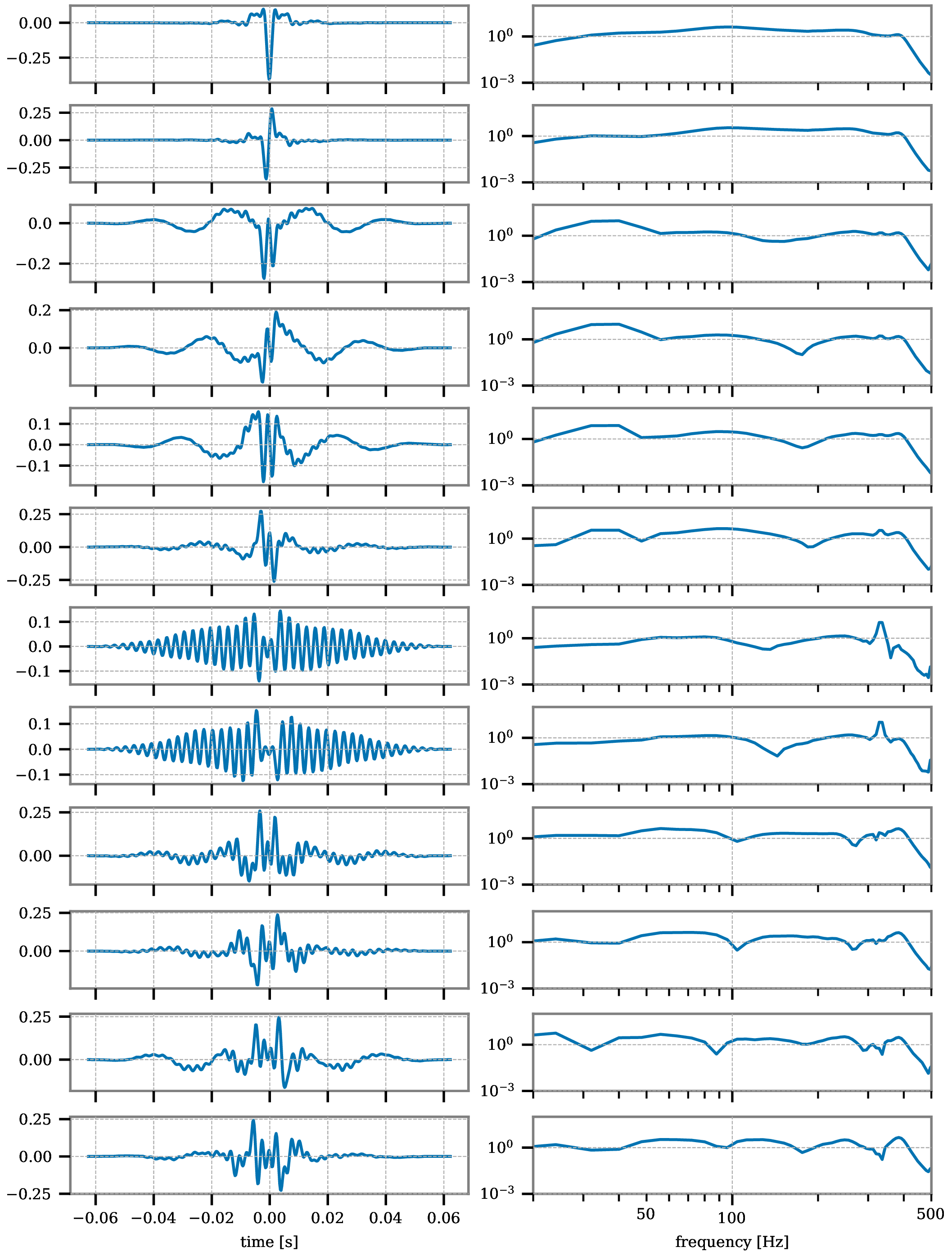}
    \caption{\label{fig:glitch_bases} The 12 components of the $V^T$ matrix used to model the glitch, in descending order of significance, in the time domain (left) and their amplitude spectral density (right).}
\end{figure*}

\cref{fig:glitch_bases} shows the components of the $V^T$ matrix, the basis functions from which the glitches are built, in both the time and frequency domain, illustrating the most important features for reconstructing a Blip glitch as given in our model. 

\subsection{Training the Flow}
\begin{figure}[h!]
    \centering
    \includegraphics[width=0.45\textwidth]{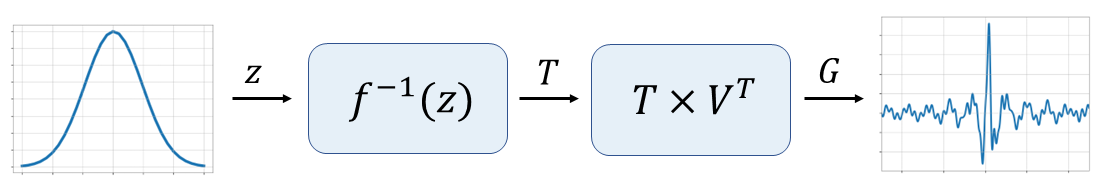}
    \caption{Illustration of glitch reconstruction. Latent parameters $z$ from a Gaussian distribution are passed through the inverse (trained) flow $f^{-1}$, and the obtained glitch amplitudes $T$ are then combined with the basis matrix $V^T$ to construct a glitch $G$. The glitch pictogram on the right is a representation of a typical Blip glitch in the time domain.} 
    \label{fig:flow_chart}
\end{figure}

We train a neural spline flow (CouplingNSF,~\cite{Durkan:2019nsq}) on the reduced glitch amplitudes $T$, using the Python library {\tt glasflow}~\cite{williams2023glasflow}.
The trained flow describes the prior distribution of the glitch parameters $T$ and can be easily sampled by drawing latent parameters $z$ from a Gaussian distribution and passing them through the inverse flow to reconstruct $T$ values corresponding to glitches. Recombining these with the basis matrix $V^T$ then gives the parametrised glitch model we use in our analysis. This process is illustrated in \cref{fig:flow_chart}.

\subsection{Applying the Glitch Model}
Once we have a trained glitch model, we can apply it with {\tt Bilby} to remove glitches from \ac{GW} data. We define a {\tt Bilby} prior class to contain the trained \ac{NF}, and a likelihood class, which can simulate the data containing the glitch along with the \ac{GW} signal. The nested sampler in {\tt Bilby} (we use {\tt Nessai}~\cite{Williams:2021qyt}) thus fits the glitch model to the glitch in the data as well as estimates the posterior of the signal.

The glitch is handled by subtracting its waveform from the data. First, the amplitudes $T$ are sampled from the saved \ac{NF} in the priors and passed to the likelihood as parameters for the glitch model. Within the likelihood, the glitch is then reconstructed using the basis matrix $V^T$, and the residual (the glitch-subtracted data) is calculated. The log-likelihood used for the signal posterior inference is thus determined from the residual, rather than the glitch-contaminated data. 

We introduce two additional unknown parameters at inference time to give the glitch model greater flexibility. A time shift is added to ensure that the glitch model can be accurately fitted to the data even if the trigger time is slightly offset, within a Gaussian prior with standard deviation 10\,ms.
We also add a scaling factor $A$ to the glitch model, with a prior $\propto A^{-1}$ within the range $10^{-3}$ to $10^3$. This allows us to adjust the overall scale of the glitch while maintaining the shape as encoded by the relative size of the $T$ parameters.

%% file: 4.analysis.tex
We performed a thorough investigation of the glitch model's performance, looking at both model selection and parameter estimation results, with and without a signal present.
In \cref{sec:glitch-only}, we test the performance of our model in identifying glitches via Bayesian model selection. We analyse interferometer data with known glitches identified by Gravity Spy, as well as simulated Gaussian noise, and compare the distribution of log Bayes factors for the glitch model and the Gaussian noise model.
In \cref{sec:signal-only}, we use the glitch model to analyse data containing signals, but no glitches, and check that the model does not misidentify signals as glitches (i.e. that it is not preferred over the signal model itself).
In \cref{sec:injections}, we test the performance of our model in the scenario where both a signal and a glitch may be present in the data. We use model selection to distinguish between signal-only, glitch-only and signal+glitch models, with a signal injected into data containing glitches. We also investigate the effect of varying the time of the glitch relative to the peak of the signal.
In \cref{sec:bias}, we investigate the bias in the recovered parameters of the injected signal when using the glitch model, and compare it to the bias in the case where no glitch model is used.
Finally, in \cref{ss:tomte} we experiment with using the Blip glitch model to remove Tomte glitches.

%% file: 5.glitch-only_fitting.tex
We first wish to establish that the glitch model is functioning well and can distinguish glitches from Gaussian data.
We investigate the recovery of glitches from the training dataset, as well as model unseen glitches. To determine if the glitch model is favoured or disfavoured for given data, we compute the Bayes factor between the glitch model and the Gaussian noise model. A log Bayes factor greater than 0 implies that the glitch model is favoured. Ideally, we would like a clean separation of results so that no Gaussian data is identified as containing a glitch, and vice versa.

\subsubsection{With known glitches}
To test the glitch model for real glitches, we select several known glitch times randomly from the Gravity Spy dataset and run the analysis on the strain data from the relevant LIGO detector around these times. This was done both for glitches present in the training data (O1) and for unseen glitches from the O2, O3a, and O3b runs. 

\begin{figure}
    \centering
    \subfigure{\includegraphics[width=0.45\textwidth]{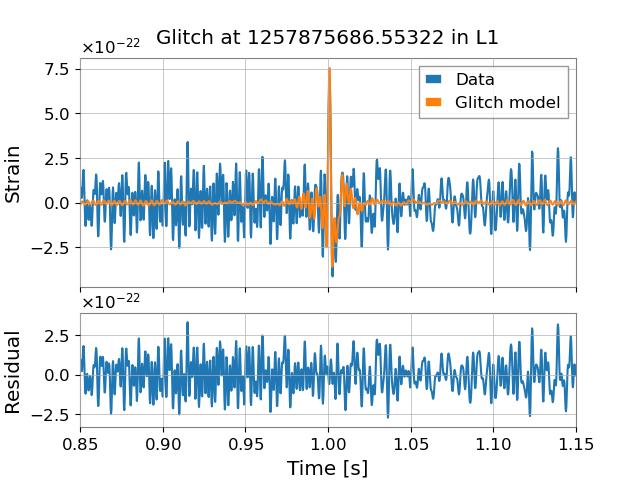}}
    \caption{(Upper panel) Plot of an O3 Blip glitch (GPS time given in title) in the LIGO-Livingston (L1) data. The filtered strain data are represented in blue, and the maximum likelihood fitted glitch model in orange. (Lower panel) The residual after subtraction of the glitch, showing no visible trace of the glitch remaining.
    \label{fig:glitchonly}}
\end{figure}

\begin{figure*}
    \centering
    \subfigure{\includegraphics[width=0.49\textwidth]{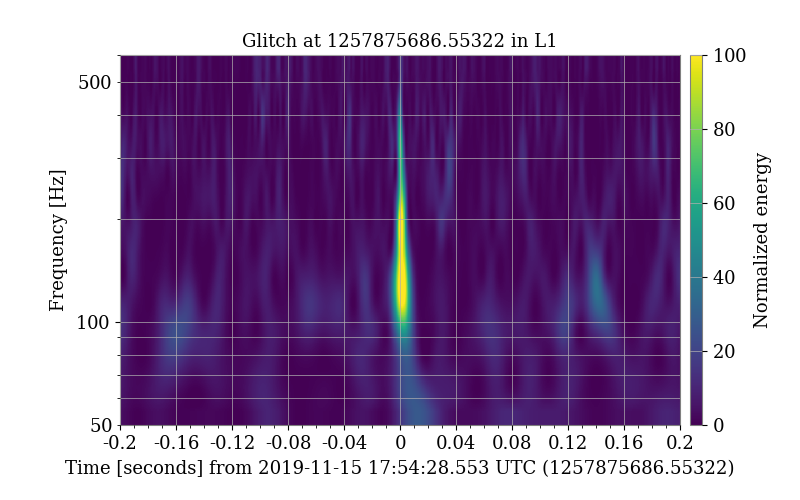}}
    \subfigure{\includegraphics[width=0.49\textwidth]{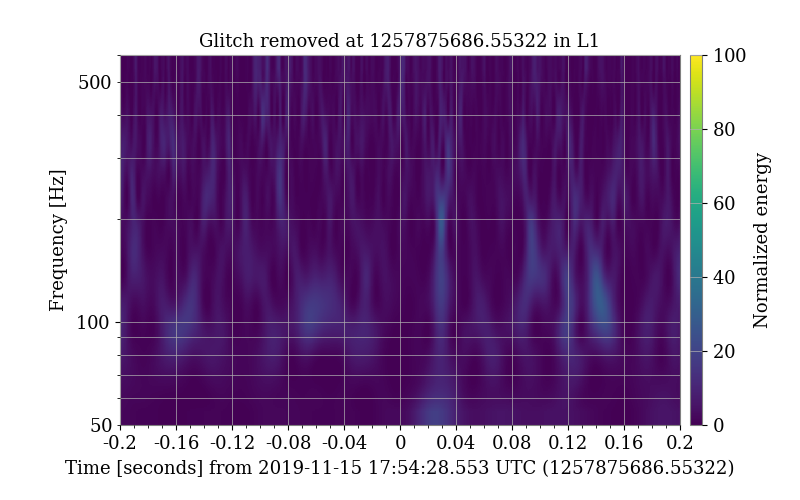}}
    \caption{Q-scan plots of the O3 Blip glitch shown in \cref{fig:glitchonly}. The plot on the left shows the data pre-glitch removal, while in the plot on the right, the glitch model has been applied to remove the glitch, leaving no visible trace.}
    \label{fig:glitchonly_qscan}
\end{figure*}

\cref{fig:glitchonly} shows an example of a glitch occurring in the LIGO-Livingston detector during the third observing run, O3. The plot shows the filtered strain data, overlaid with the maximum-likelihood glitch model fitted to the data by {\tt Bilby} (the full result also contains uncertainty on the glitch parameters). The residual is plotted underneath and used as a first reference to show that the glitch has been successfully removed. 
Furthermore, we can also use time-frequency-energy spectrogram plots (so-called Q-scans) to verify if a glitch has been removed fully. In \cref{fig:glitchonly_qscan}, the Q-scan of the same example glitch from \cref{fig:glitchonly} is shown, before and after glitch removal. The Blip glitch is clearly visible in the centre of the left-hand plot, while no visible trace of it is left after subtraction in the plot on the right-hand side.
While not quantified, this demonstrates the successful removal of all of the glitch power in a view familiar to \ac{GW} analysts.

To quantitatively determine if the glitch model is favoured or disfavoured for given data, we investigate the log Bayes factors for the glitch model. To summarise the results, we provide a histogram of all log Bayes factors obtained from fitting the glitch model to glitches in O1 and O3 data, respectively, in \cref{fig:background_foreground_hist_O1} and \cref{fig:background_foreground_hist_O3}. Almost all of the log Bayes factors are greater than zero, demonstrating that the glitch model is favoured over a Gaussian noise model, and thus showing that there is a glitch in the data, well described by the glitch model. There is one (of 786) glitch for O3 where the log Bayes factor is less than zero \changed{($-0.22$)}, thus indicating that a Gaussian noise model is \changed{slightly} preferred over the glitch model for this glitch. Investigating this further, we note that the outlier glitch appears louder and broader in the time-frequency spectrogram than most Blip glitches, and it is possible that this is the reason for the glitch model not being preferred \changed{as is looks different from the Blip glitches the model is trained on} (see \cref{ss:tomte} for further discussion).

\subsubsection{On Gaussian noise}
\begin{figure*}
    \centering
    \subfigure{\includegraphics[width=0.45\textwidth]{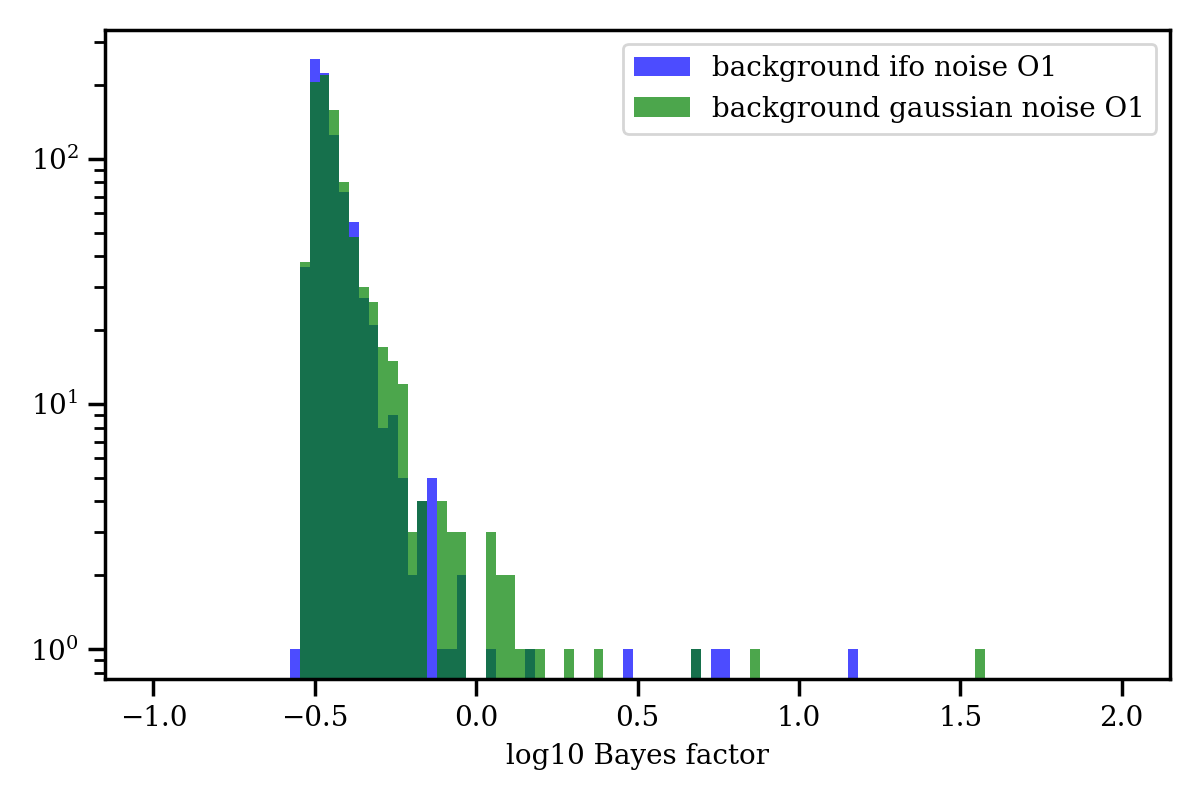}}
    \subfigure{\includegraphics[width=0.45\textwidth]{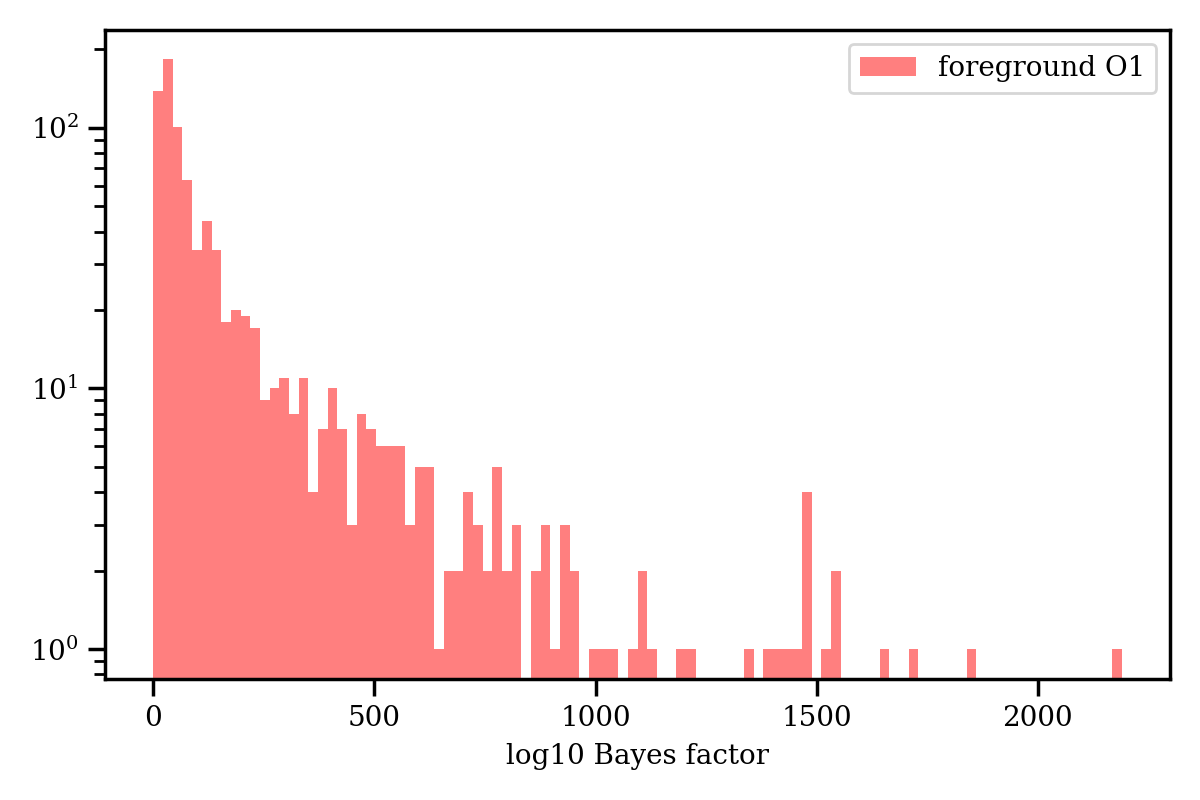}}
    \caption{Background: Histogram of log Bayes factors obtained from applying the glitch model to glitch-free O1 interferometer data and Gaussian noise data. The log Bayes factors are mostly below zero, except for 7 of 860 runs on detector noise and 15 of 881 runs on Gaussian noise. Foreground: Histogram of log Bayes factors from applying the glitch model to glitches in the O1 strain data. The log Bayes factors are all above zero.}
    \label{fig:background_foreground_hist_O1}
\end{figure*}

\begin{figure*}
    \centering
    \subfigure{\includegraphics[width=0.45\textwidth]{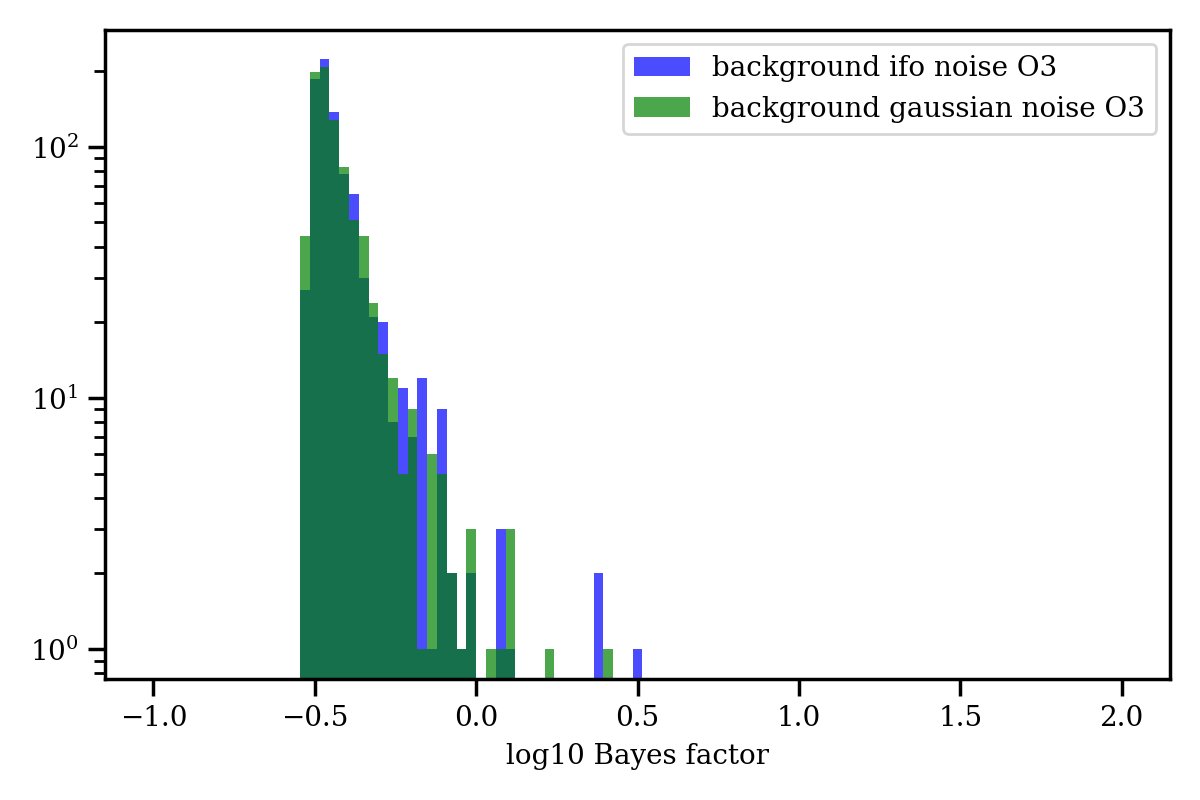}}
    \subfigure{\includegraphics[width=0.45\textwidth]{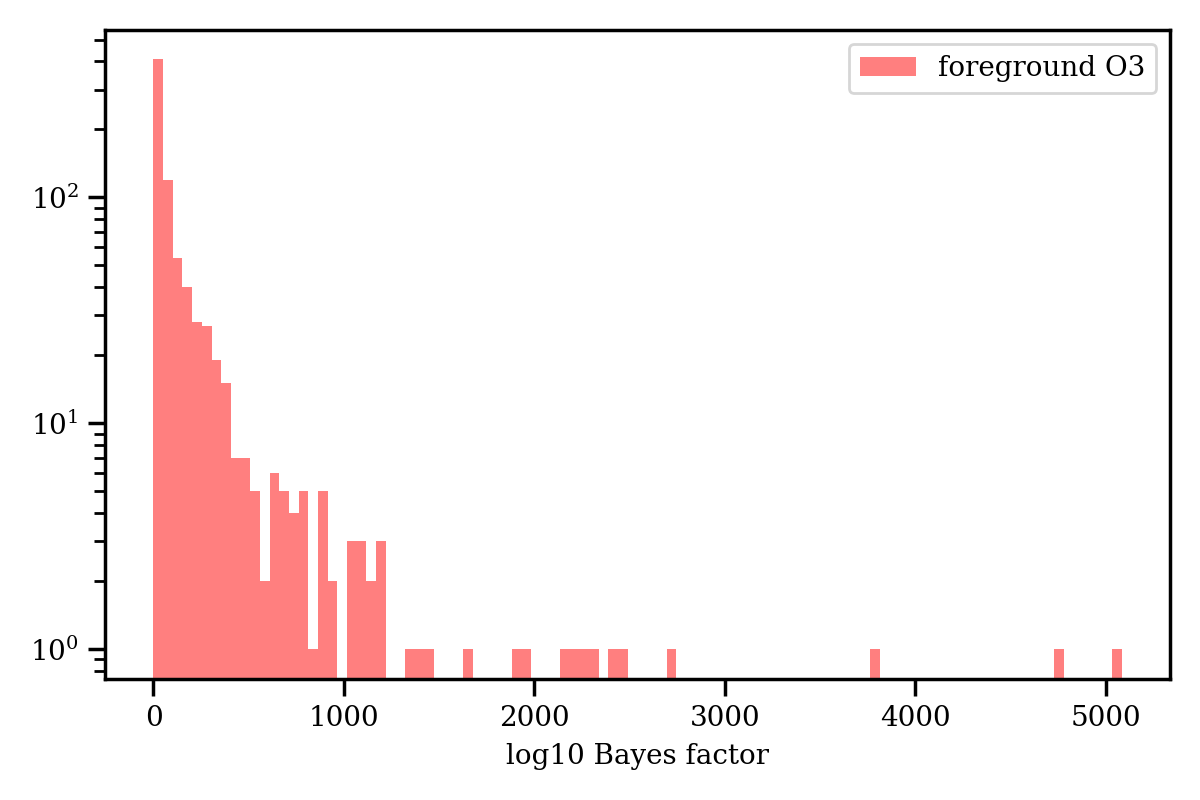}}
    \caption{Background: Histogram of Bayes factors obtained from applying the glitch model to glitch-free O3 interferometer data and PSD noise data. The Bayes factors are mostly below zero, implying that the glitch model is disfavoured, except for 7 of 849 runs each on detector and Gaussian noise. Foreground: Histogram of log Bayes factors from applying the glitch model to glitches in the O3 strain data. The log Bayes factors are all above zero, except for one of the 786 glitches (with a log Bayes factor value $-0.22$).}
    \label{fig:background_foreground_hist_O3}
\end{figure*}

To check that the glitch model is disfavoured when applied to data not containing a glitch, we apply it to Gaussian noise generated according to a known power spectral density (PSD), as well as to LIGO strain data at times when no glitch (or signal) is present (according to Gravity Spy, using a Omicron SNR threshold of 7.5~\cite{Glanzer:2022avx}). The test runs returned log Bayes factors of values less than zero for the majority of cases. The mostly negative log Bayes factors demonstrate that the glitch model is disfavoured as expected when no glitch is present, and hence will not give false alarms. The resulting background Bayes factor histograms are shown in \cref{fig:background_foreground_hist_O1} and \cref{fig:background_foreground_hist_O3}. Comparing the background histograms for O1 and O3, we note that the distributions look very similar between runs and between the Gaussian noise and the detector data. We find that 7 of 860 runs on detector noise and 15 of 881 runs on Gaussian noise have log Bayes factors greater than zero for O1, while there are 7 outliers each for O3 (out of 849 runs each on detector and Gaussian noise). From a brief investigation of the outliers, we note that visually, there appears to be a small amount of excess power in the strain data plots. It is worth highlighting that almost all of the outliers have log Bayes factors smaller than 0.5, indicating that the evidence in favour of the glitch model is ``not worth more than a bare mention'' according to the commonly used table by \citet{Kass1995BF}. Furthermore, none of the outliers have log Bayes factors above 2, which by the above classification indicates the evidence is not decisive. The good separation between our background and foreground Bayes factor distributions indicates that naive Bayesian model selection with equal prior odds for each hypothesis would result in a posterior odds ratio that serves as a reasonable detection statistic. A more sophisticated analysis could take into account the rate of glitches (and later signals) or target an expected false alarm probability by adjusting the prior probability for each hypothesis considered, but we find this unnecessary in the present work~\cite{Veitch:2008wd}.

Overall, these results indicate that the model is able to cleanly separate the populations of glitches and non-glitchy data with little overlap in the Bayes factor statistic.

%% file: 6.signal-only.tex
Next, we test the glitch model in the presence of an injected signal of a binary black hole merger. It is important to first confirm that the glitch model does not affect the results and that the glitch model is not favoured over the signal model if no glitch is present in the data. 

\begin{figure}[h!]
    \centering
    \subfigure{\includegraphics[width=0.49\textwidth]{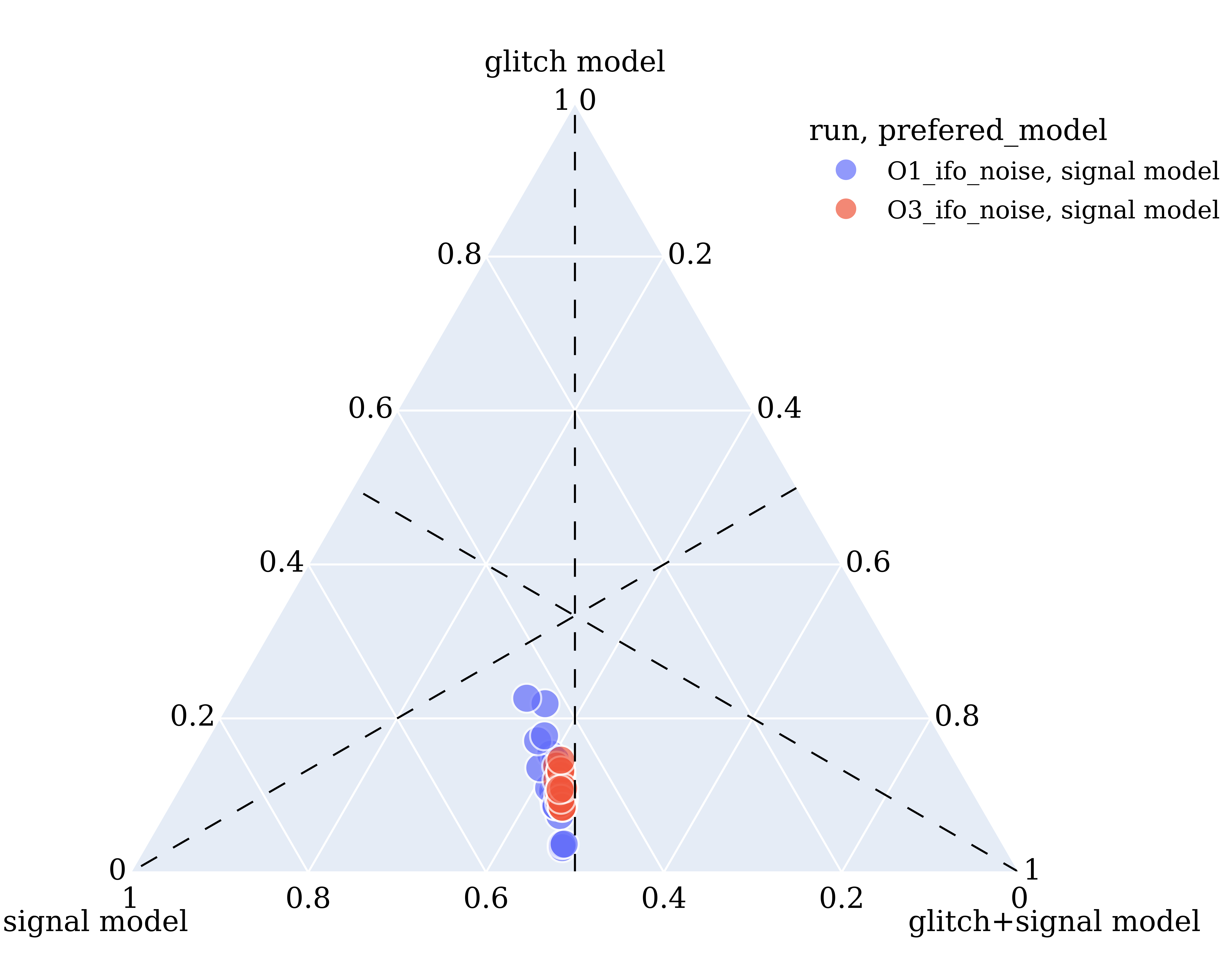}}
    \caption{The ternary plot shows the model comparison results, as calculated on detector data with an injected signal and no glitches, when applying both the glitch and signal model, and when applying only the glitch and only the signal model, respectively. The dashed lines represent the boundaries between preferred models. Events are coloured by the observing run the data originate from, and the shapes are determined by which model is preferred by the log Bayes factor for each test glitch. }
    \label{fig:logBF_ternary_noise}
\end{figure}

To show that the glitch model is not favoured, we inject a signal into noisy data from the interferometers at times without glitches and apply both the glitch and signal models.~\cref{fig:logBF_ternary_noise} shows a ternary plot of the log Bayes factors for the three cases: glitch and signal model, glitch model only, and signal model only. The dashed lines indicate the boundaries between preferred models. We repeat the experiment for detector noise from observing runs O1 and O3. In both cases, we find that the model containing the signal only is preferred, and we have thus shown that the glitch model is not preferred when there is no glitch in the data. We find that in all cases the signal-only model is quite strongly preferred over the glitch-only model, and that it is also preferred less strongly over the glitch+signal model. As expected, the glitch+signal model is only slightly disfavoured, as this glitch model can contain low amplitude glitches consistent with the data, and the lower Bayes factor comes from the Occam factor penalising the model for unnecessary complexity.

%% file: 7.injection-tests.tex
To test our model in the presence of a signal and a glitch, we now inject \ac{GW} signals into glitch-contaminated data. By comparing the injected values to the obtained posteriors, we can determine if the results are improved by applying our glitch model to remove the glitch prior to the parameter evaluation. For the comparison, 6 of the 15 parameters typically used to describe a binary black hole merger were chosen as the parameters of interest: mass ratio $q$, chirp-mass $\mathcal{M}$, inclination $\theta_{JN}$, luminosity distance, right ascension, and declination. \changed{We chose to only present theses parameters for clarity and because these are the most interesting and important for how they affect the signal model.}

\begin{figure}[h!]
    \centering
    \subfigure{\includegraphics[width=0.5\textwidth]{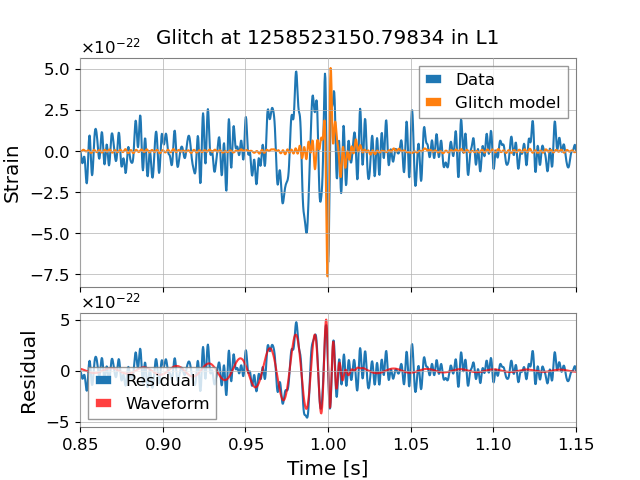}}
    \caption{Plot of data containing a glitch during an injected binary black hole signal, together with the fitted glitch model. The residual in the lower panel is overlaid by the waveform model for the injected signal.\label{fig:injection}}
\end{figure}

In \cref{fig:injection} we show an example of the glitch model being fitted to strain data containing both a glitch and an injected signal. The waveform of a source at a luminosity distance of 750 Mpc was injected with the merger happening at the time of the glitch, thus overlapping the signal waveform with the glitch. The plot clearly demonstrates how, once the glitch is removed from the data, the signal becomes clearer. The waveform of the injected signal is plotted together with the residual to highlight this further. 

\begin{figure}[h!]
    \centering
    \subfigure{\includegraphics[width=0.49\textwidth]{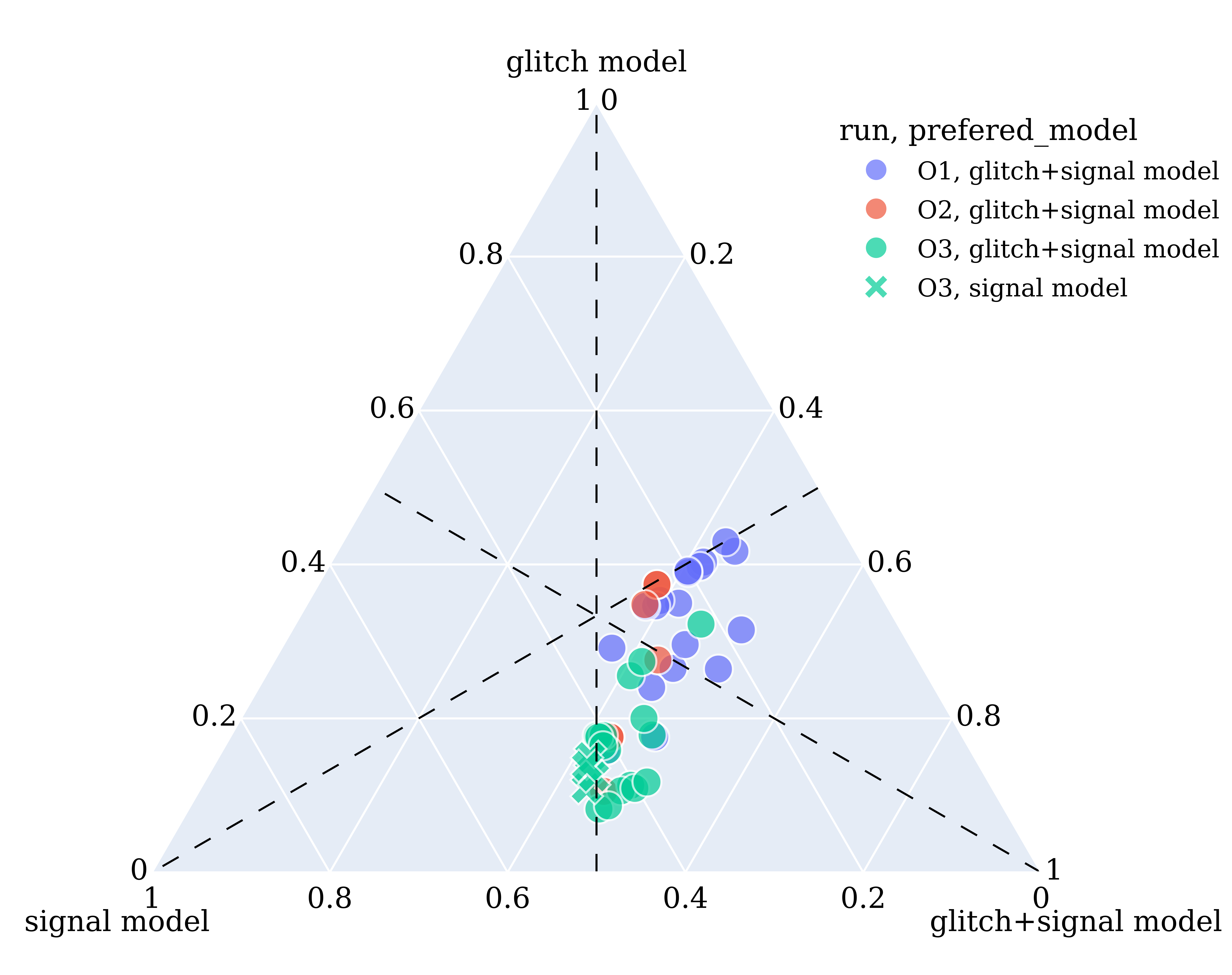}}
    \caption{The ternary plot shows the log Bayes factors (as calculated on data containing both an injected signal and a glitch) when applying both the glitch and signal model, and when applying only the glitch and only the signal model respectively. The dashed lines represent the boundaries between preferred models. Events are coloured by the observing run the glitchy data originate from, and the shapes are determined by which model is preferred by the log Bayes factor for each test glitch.\label{fig:logBF_ternary}}
\end{figure}

We can again make use of the Bayes factors to compare models and determine how well the glitch model is doing. Running on data containing both a glitch and an injected signal, we compare the case where we have both the glitch and signal model to when there is only the glitch and only the signal model. We expect larger Bayes factors for the model preferred by the data. In \cref{fig:logBF_ternary}, the log Bayes factors for these three cases are plotted against each other in a ternary plot. The figure contains example glitches from all three observing runs. We observe that the case where both models are applied is preferred for the majority of our test glitches. However, in four of the 25 O3 test glitches, the signal-only model is slightly favoured. 

A possible explanation could be that the background is not well measured. For example, if the noise is non-stationary around the time of the glitch, the detector output varies throughout the duration of the data analysed, affecting the results.
Alternatively, this could be explained by these glitches all being very weak and having a slightly less distinct shape than typical for Blip glitches. Thus, applying the glitch model does not improve the signal analysis sufficiently for this model to be favoured. Both the q-scans and residual plots only show a minimal difference before and after the glitch has been removed. Looking at the filtered time data, these glitches appear to have an amplitude comparable to the injected signal and could, by eye, be mistaken for part of the signal. The q-scans before glitch-removal also do not appear to have a visible glitch in the signal, and the power only shifts minimally when the glitch is removed. From these observations, it is comprehensible why the signal-only model is slightly favoured over the glitch-and-signal model case. However, investigating these glitches further, we find that the signal posteriors still appear to be improved when the glitch model is applied. Studying the bias for these four glitches, we note that for all parameters, the bias is comparable between the glitch-contaminated and glitch-removed cases. Hence, removing the glitch does not reduce the bias much, if at all, for these cases.

\begin{figure}[h!]
    \centering
    \subfigure{\includegraphics[width=0.45\textwidth]{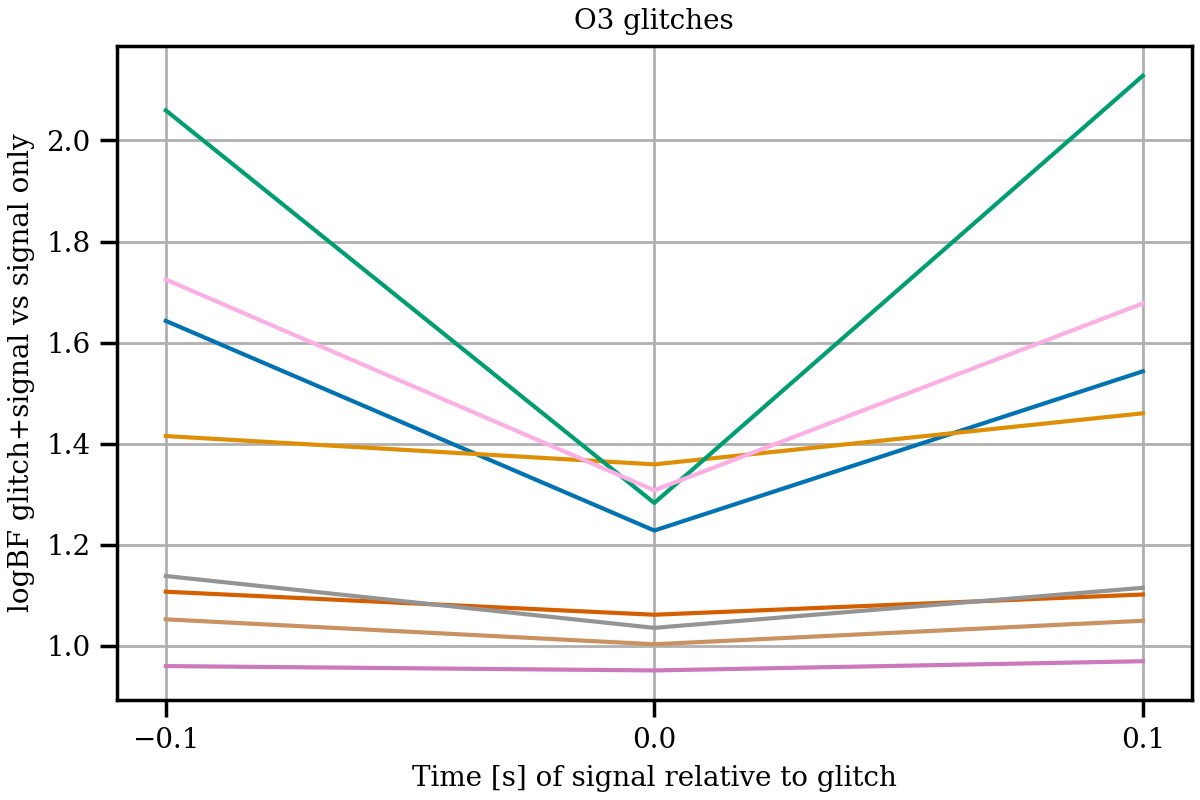}}
    \caption{The plot shows how the Bayesian model preference changes depending on if the signal is injected before, during or after the glitch. The $x$-axis represents the three relative times when the signal is injected, and the $y$-axis shows the ratio of log Bayes factors for the glitch+signal model versus the signal model only. Each line in the plot represents a specific test glitch. The uncertainty on the log Bayes factors is around 0.1 for all test glitches, and thus not shown in the plot.\label{fig:logBF_scatter_lags}}
\end{figure}

Furthermore, we briefly investigate the effect of the glitch being placed before, during or after the signal merger. So far, all experiments have considered a glitch overlapping with the merger. We now perform the model selection experiment, as above, for signals injected 0.1 seconds before and after the glitch. Running on data containing both a glitch and an injected signal, we compare the case where we have the glitch and signal model to when there is only the signal model. In \cref{fig:logBF_scatter_lags}, the log Bayes factors between these two cases are plotted on the $y$-axis for a few of the test glitches from the third observing runs. The $x$-axis shows the three signal injection times considered: before, during and after the glitch. 

We observe that the log Bayes factors vary for the same glitch, depending on where the signal is injected in relation to the glitch. The log Bayes factor has similar values when the signal is injected before and after the glitch, indicating that the model preference is almost the same between these two cases. Meanwhile, when the signal is injected during the glitch, compared to the before and after cases, the glitch+signal model is somewhat less favoured for most glitches. This is because it is easier for the signal model to absorb some of the glitch power if the signal is injected during the glitch, thus reducing the need for the glitch model. Due to the signal model being time-constrained, it is difficult for additional glitch power to be absorbed when the glitch and signal do not overlap.

Furthermore, all three cases for each glitch are either above or below one, which is the limit separating which model is preferred. Thus we find the four glitches for which the signal-only model is favoured in \cref{fig:logBF_ternary}, have values below one for all three cases of the signal being injected before, during and after the glitch (only one of these examples is shown in \cref{fig:logBF_scatter_lags}). For no other test glitches, of any lag, is the signal-only model preferred over the glitch+signal model. Hence, the quietness and properties of the glitch itself must be the reason for the glitch+signal model not being preferred, rather than how the glitch affects the signal.

%% file: 8.bias-tests.tex
As a bias test, we inject a signal into glitch-contaminated data and analyse the signal before and after removing the glitch. We can then compare the inference of the posterior values of the injected signal. 

To quantify the bias, we compute the `standard accuracy', which is defined for parameter $x$ as
\begin{equation}
    \Sigma_{x} = \frac{|x_\mathrm{maxL} - x_\mathrm{true}|}{\sigma_x},
\end{equation}
where $x_\mathrm{maxL}$ is the maximum likelihood posterior value of $x$, $x_\mathrm{true}$ is the injected value, and $\sigma_x$ is the standard deviation of the posterior of $x$. For an unbiased estimate of the posterior, we would expect the standard accuracy to be around $1$ due to statistical fluctuations only.

\subsubsection{On Blip glitches} 
\begin{figure*}
    \centering
    \subfigure[Posterior for glitch-contaminated signal]{\includegraphics[width=0.45\textwidth]{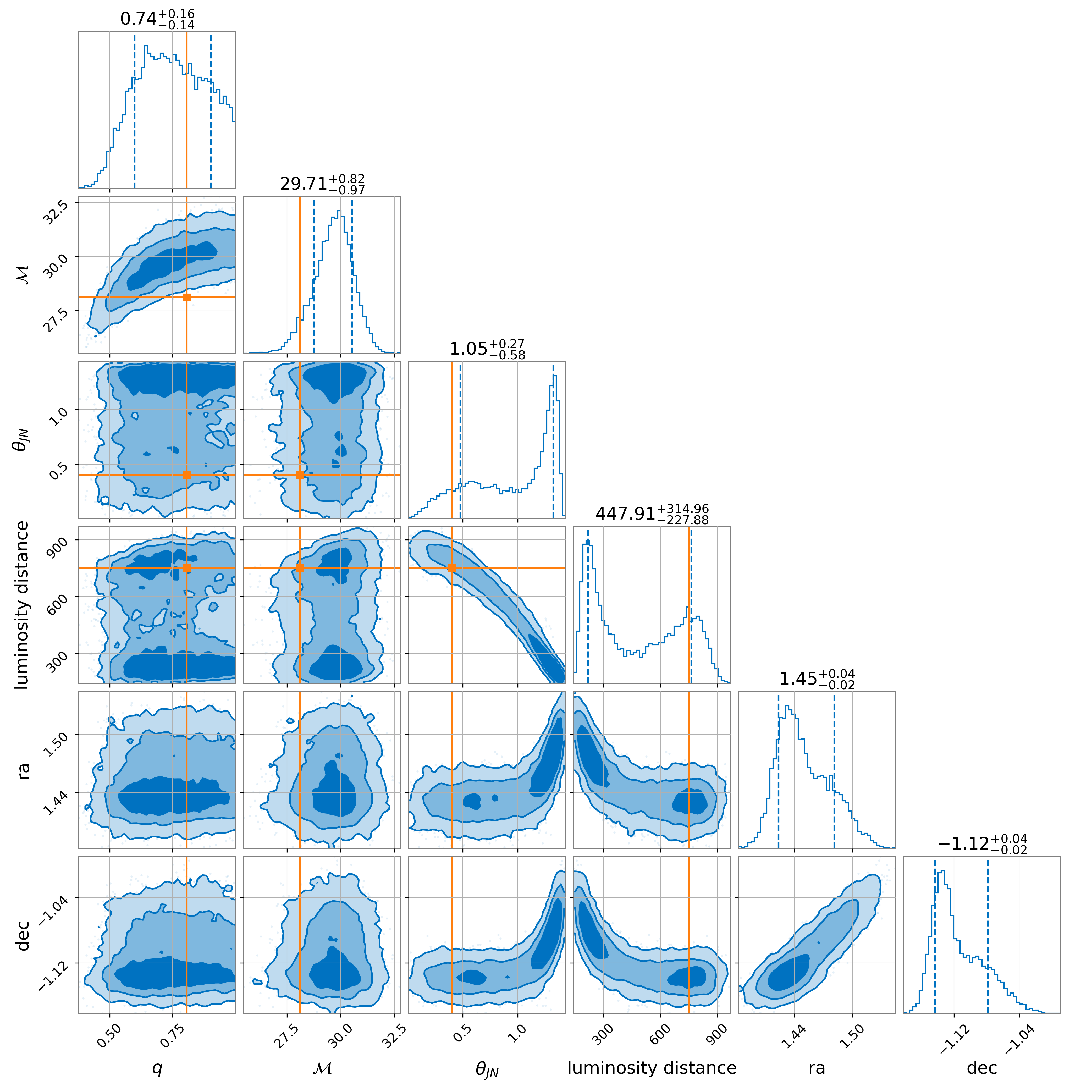}}
    \subfigure[Posterior for glitch-removed signal]{\includegraphics[width=0.45\textwidth]{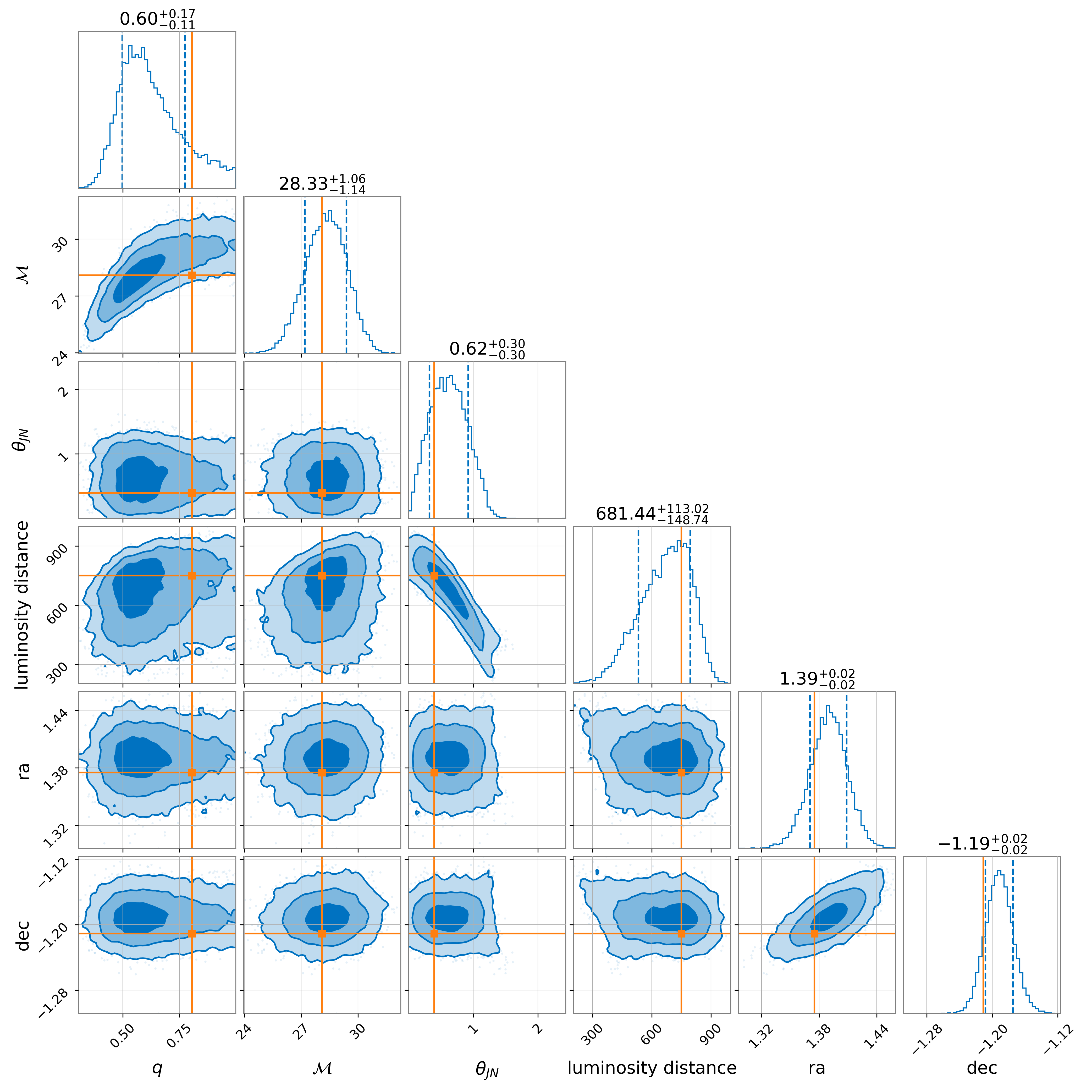}}
    \caption{Corner plots of signal posteriors, inferred before and after the glitch was removed from the data, respectively. For clarity, only mass ratio $q$, chirp-mass $\mathcal{M}$, inclination $\theta_{JN}$, luminosity distance, right ascension, and declination are shown. The blue contours show the 16\% and 84\% percentiles of the posterior distributions, and the cross-hairs in each subplot indicate the injected values. This is the posterior for the same glitch also shown in figure \cref{fig:injection}.\label{fig:posteriors}}
\end{figure*}

The two corner plots in \cref{fig:posteriors} show the signal posteriors as inferred from data with or without the glitch being removed, respectively. The corner plots in \cref{fig:posteriors} demonstrate how the inference of the signal posterior improves significantly if the glitch is removed from the data prior to analysis. 

\begin{figure}[h!]
    \centering
    \subfigure{\includegraphics[width=0.5\textwidth]{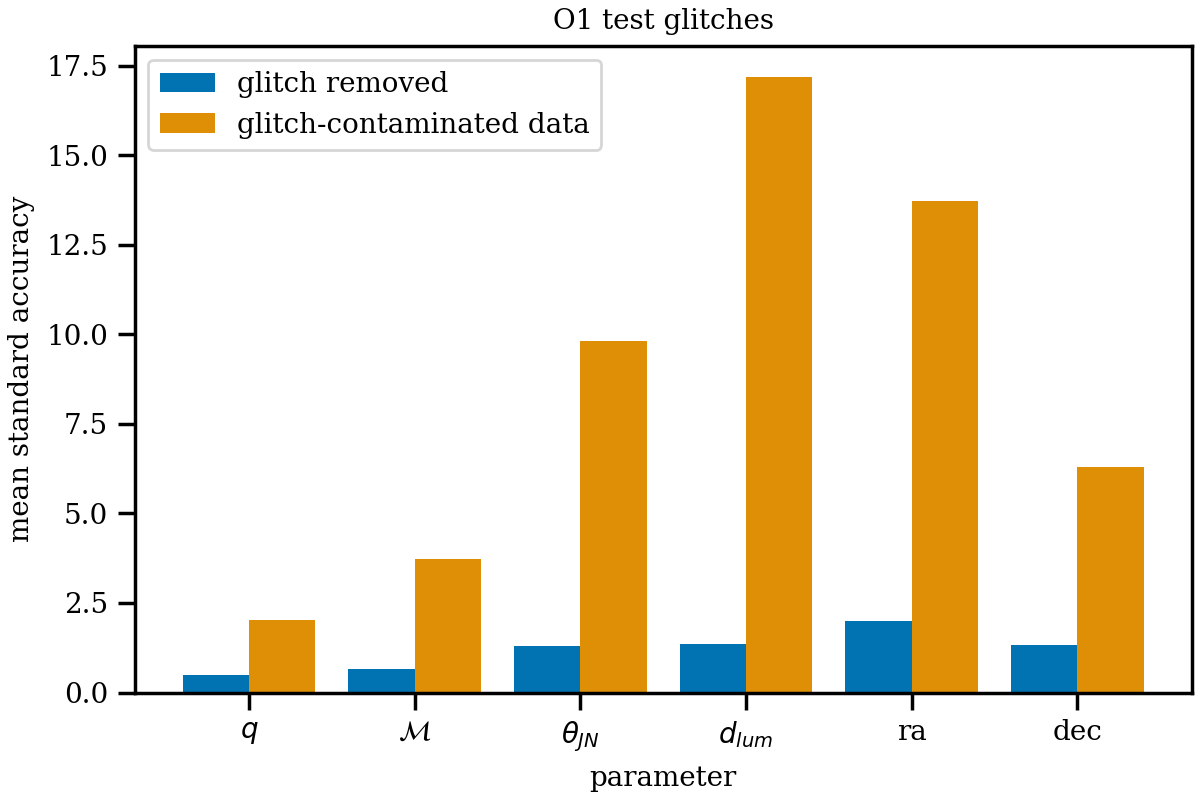}}
    \caption{Mean standard accuracy over 25 O1 test glitches, for a few parameters: mass ratio $q$, chirp-mass $\mathcal{M}$, inclination $\theta_{JN}$, luminosity distance $d_{lum}$, right ascension, and declination.}\label{fig:standard_accuraccy_O1}
\end{figure}

\begin{figure}[h!]
    \centering
    \subfigure{\includegraphics[width=0.5\textwidth]{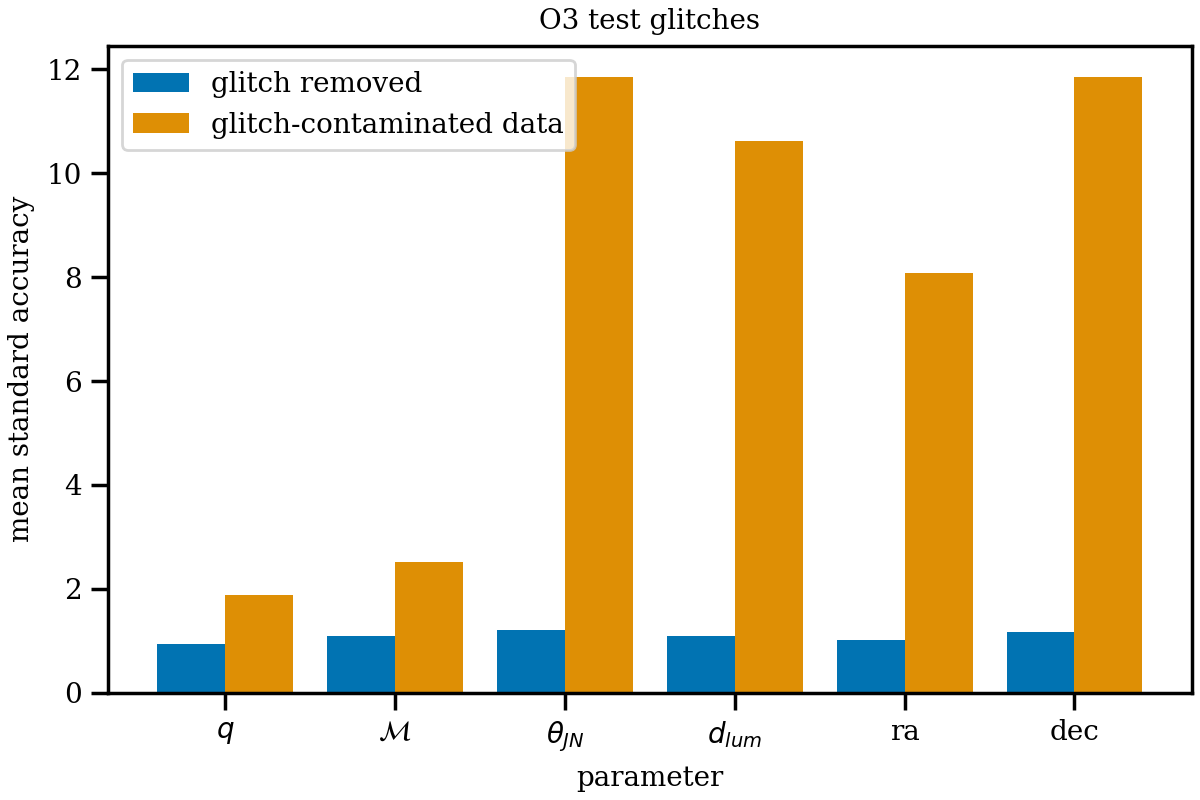}}
    \caption{Mean standard accuracy over 25 O3 test glitches, for a few parameters: mass ratio $q$, chirp-mass $\mathcal{M}$, inclination $\theta_{JN}$, luminosity distance $d_{lum}$, right ascension, and declination.}\label{fig:standard_accuraccy_O3}
\end{figure}

The mean of the standard accuracy of a few parameters for 25 O1 test glitches are shown in \cref{fig:standard_accuraccy_O1}, and for O3 in \cref{fig:standard_accuraccy_O3}. The plots show a lower mean standard accuracy for the case where the glitch has been removed from the data for all parameters. We can thus conclude that applying the glitch model improves the signal parameter estimation by reducing the bias. It is also interesting to note that the mass parameters seem to be less affected by the glitch-contamination than the other parameters shown in the plots. This is likely because the presence of a glitch in only one detector cannot create as large a deviation in the phase as it can in the relative amplitude between detectors that more strongly informs the extrinsic parameters.

In the above analysis, the signal is injected during the glitch. We also investigate how the bias standard accuracy changes if the signal is injected before or after the glitch. We find that the mean bias standard accuracy for the glitch-contaminated data is significantly higher (for all parameters except the mass ratio $q$) when the signal is injected during the glitch compared to before or after. There is little to no difference between the latter two cases. For the glitch-removed data, the bias standard accuracy is around one for all parameters, regardless of where in time the signal is injected. This shows that although the bias is higher before glitch removal when the signal coincides with the glitch, it is still successfully reduced when considering the glitch-removed case. 

The results are similar for both the O1 and O3 test glitches, although the variations between injections before or after the glitch are slightly larger in O1, with the signal being injected before the glitch having slightly larger biases for most parameters. 

\subsection{On Tomte glitches}\label{ss:tomte}
To explore how our glitch model, trained on Blip glitches, behaves when applied to a different type of glitch, we perform the same tests for Tomte glitches. Tomte glitches have a similar morphology to Blip glitches and are hence interesting candidates for this experiment. 

\begin{figure}
    \centering
    \subfigure{\includegraphics[width=0.45\textwidth]{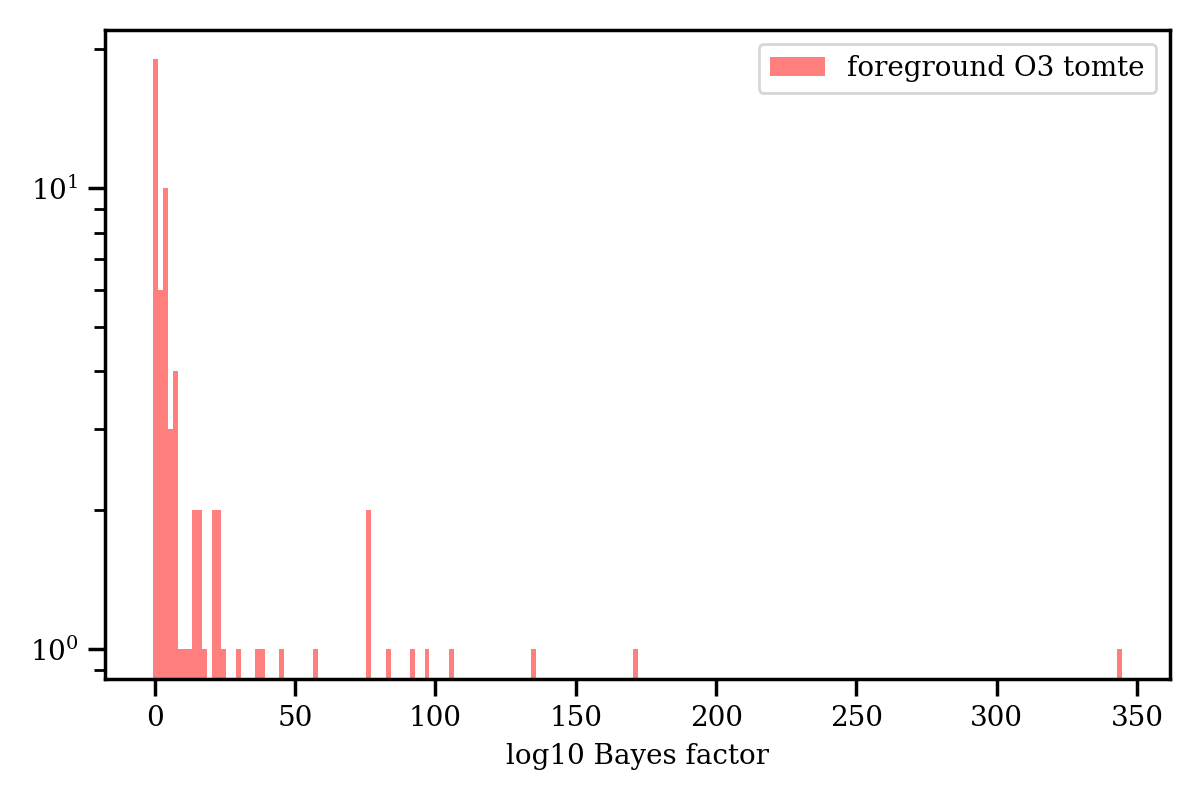}}
    \caption{Foreground histogram of log Bayes factors from applying the glitch model (trained on Blip glitches) to Tomte glitches in the O3 strain data. The log Bayes factors are mostly above zero, but 15 of the 69 test glitches have values below zero.}\label{fig:foreground_hist_tomte}
\end{figure}

Firstly, we plot a foreground histogram of log Bayes factors for 69 Tomte glitches found in O3 data, see \cref{fig:foreground_hist_tomte}. Similarly to the case with Blip glitches in the data, we find that the glitch model is preferred over the noise model for all except 15 of the 69 glitches. This demonstrates that although the model is trained on Blip glitches, it can also successfully remove Tomte glitches from the data in certain cases, but the success rate is significantly lower than for removing Blip glitches. 

\begin{figure}[h!]
    \centering
    \subfigure{\includegraphics[width=0.45\textwidth]{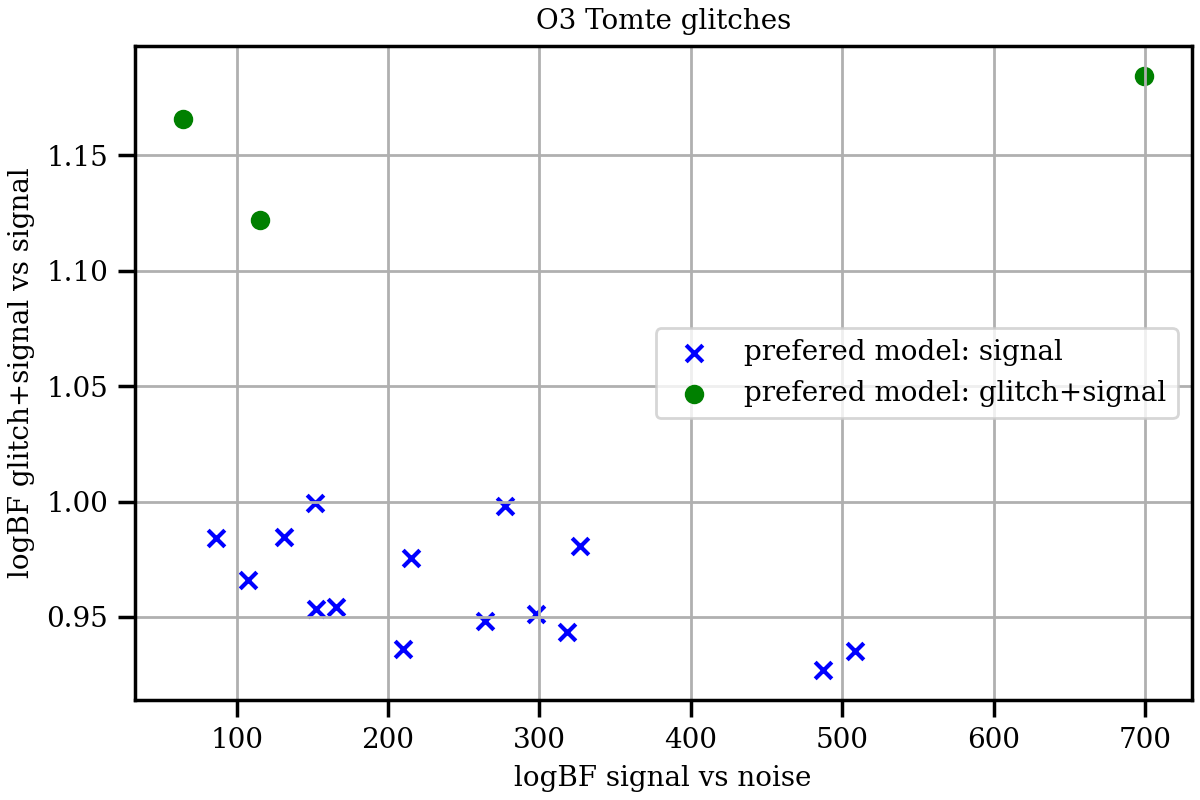}}
    \caption{The plot shows the log Bayes factors when applying the (Blip-trained) glitch and signal model versus the signal model only to O3 data containing Tomte glitches and signal injections. The x-axis shows the signal-only model vs noise model log Bayes factors, and the y-axis shows the ratio between the log Bayes factors of the glitch+signal model and the signal-only model. The uncertainty on the log Bayes factors is around 0.1 for all test glitches, and thus not shown in the plot. }\label{fig:logBF_scatter_tomte}
\end{figure}

Secondly, we inject signals in data contaminated by Tomte glitches and compare the analyses when the glitch model is applied and not. The resulting log Bayes factor scatter plot is shown in \cref{fig:logBF_scatter_tomte}. We note that almost all points are below one, indicating that including the (Blip-trained) glitch model in the analysis is not preferred. However, there are three (of 20) points above one, likely due to these Tomte glitches being particularly similar to Blip glitches. Some of the glitches are very close to one, showing that the signal-only model is only slightly preferred by the data. 

\begin{figure}[h!]
    \centering
    \subfigure{\includegraphics[width=0.45\textwidth]{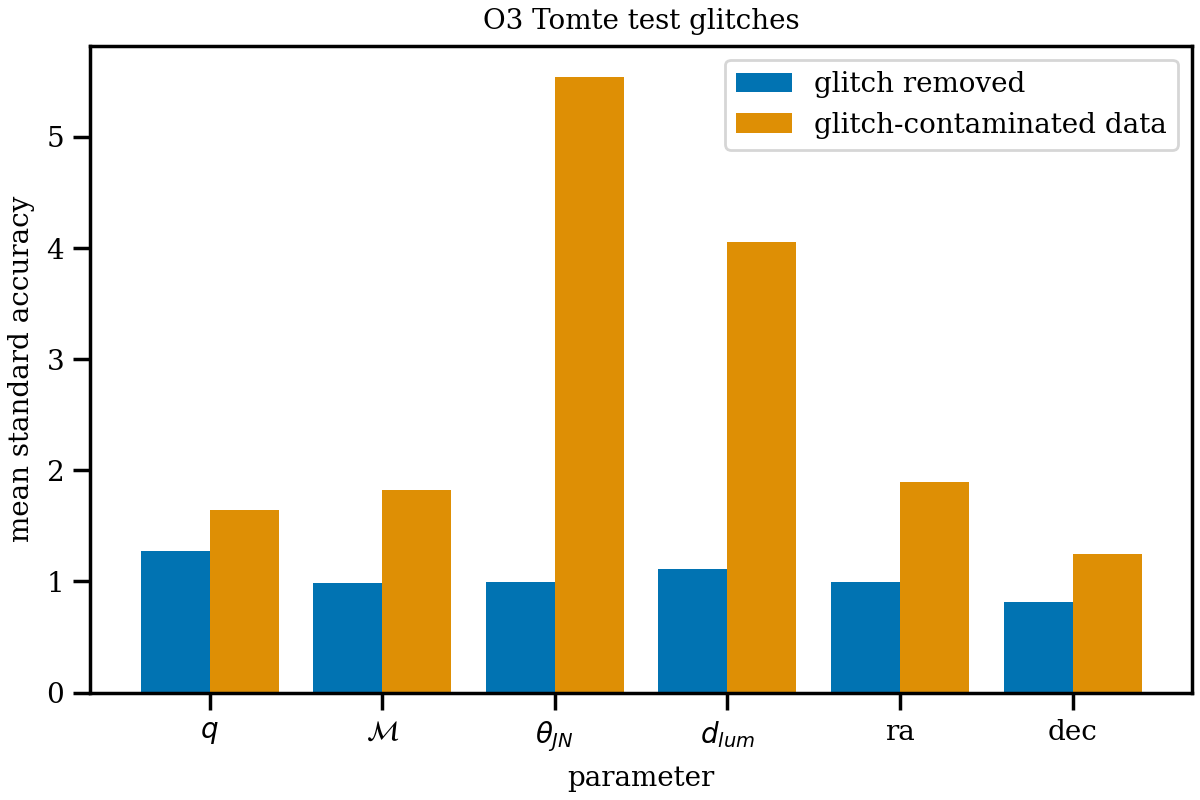}}
    \caption{Mean standard accuracy for 20 O3 Tomte test glitches, for signal parameters mass ratio $q$, chirp-mass $\mathcal{M}$, inclination $\theta_{JN}$, luminosity distance $d_{lum}$, right ascension, and declination.}\label{fig:standard_accuraccy_O3_tomte}
\end{figure}

Lastly, we repeat the standard accuracy calculations for the signals injected into data containing Tomte glitches. The results are shown in \cref{fig:standard_accuraccy_O3_tomte}. The plot shows that all parameters have a lower mean standard accuracy when the glitch has been removed, even though the glitch model was trained on a different type of glitch than what is being removed. We thus show that removing the glitch, even with an incorrect model, reduces the bias in the signal. Remembering \cref{fig:standard_accuraccy_O1} and \cref{fig:standard_accuraccy_O3}, however, the bias was reduced significantly more between the glitch-contaminated and glitch-removed cases for the Blip glitches than for the Tomte glitches in \cref{fig:standard_accuraccy_O3_tomte}. The standard accuracy for the glitch-removed cases is, on average, around 1 for both Blip and Tomte glitches. However, the bias goes up to an average of 5 for Tomte and up to about 17/15 for Blip glitches for the glitch-contaminated cases.

%% file: 10.summary.tex
In this work we have shown that modelling Blip glitches using normalising flows can successfully remove glitches from the data, thus reducing bias in the \ac{GW} signal parameter estimation. We have shown that our glitch model successfully removes the glitches we have applied it to, without significantly affecting signals also present in the data. 

Firstly, we explored the case of glitch removal only and found that our glitch model reliably removes all test glitches. Applying the model to a variety of Blip glitches in both O1 and O3 data, we obtained 0\% and 0.1\% false dismissal rates, respectively. Similarly, applying the glitch model to non-glitchy detector noise and Gaussian noise data, we find false alarm rates of 0.8\% and 1.7\%, respectively, for O1 and 0.8\% for both backgrounds in O3. Thus, we have shown that our model can cleanly separate the populations of Gaussian noise and glitches with little overlap. 
Secondly, we injected signals onto real glitches and tested the performance of our glitch model for joint signal and glitch inference. We found that the joint signal+glitch model was favoured for the majority of our test glitches when applying Bayesian model selection to data containing a glitch and a signal. Meanwhile, applying the glitch model to data containing a signal but no glitch, we found that the signal only model is preferred over the glitch+signal model for all test glitches. 
Finally, we studied the bias in the signal posterior parameters induced by the presence of a glitch. We found that removing the glitch with our model significantly reduces the bias and improves the parameter estimation of the signal. 

To train the model, we make use of the Blip glitches identified in Gravity Spy's O1 training data to train our glitch model. Hence, the model is only knowledgeable about these types of glitches as they appeared during O1. If glitch models of this kind are to be applied in future observing runs, and to other glitches, it might be relevant to train models on more recent data as the detectors are continuously improved. Some glitch classes change, and new ones might appear between detector upgrades~\cite{Soni:2021cjy, Glanzer:2022avx, LIGO:2024kkz}. For optimal performance, the glitches in the training data should be as similar as possible to the glitches that the model is then applied to remove. 

\changed{The computational cost of the analysis increases when including the glitch model.
We run the code on four CPU cores using the nessai sampler. On average, the sampling time for fitting only the glitch model to data containing a glitch is approximately $1$ hour. Fitting the glitch-only, signal-only and glitch+signal models to data containing both a glitch and a signal we get median sampling wall times (over our 25 test cases) of $6$, $0.5$, and $30$ hours respectively. There is scope for further optimisation of our likelihood function and normalising flow prior within bilby, which we will investigate in the future. However the run time is not prohibitive for an offline analysis. Furthermore, training a \ac{NF} to create the glitch model only takes a few minutes.}

This paper has focused on exploring the performance of a model trained on Blip glitches, but the methodology is applicable to any glitch class, or indeed any temporally isolated feature in the data. \changed{A similar approach may, for example, be useful for signal model discovery where no clean mapping to physical parameters is known, such as e.g. burst signals from supernovae.} We briefly investigate how our method performs for other, more complex, glitch types; we repeat the training of the model as before, now for Koi Fish and Power Line glitches, and then run the glitch-only analysis as before. We find that the glitch model removes the glitches as expected and returns log Bayes factors favouring the glitch model. We do not repeat all the extensive tests for these glitch classes, but leave it to future work to explore these and other classes fully. However, our brief investigation still shows the potential of our method being applicable to other glitch classes as well. 

From the studies on removing Tomte glitches with a Blip trained model in \cref{ss:tomte}, we observe that although the model removes the glitch, the model selection results are biased. The foreground analysis on Tomte glitches shows a 22\% false alarm rate. Injecting signals onto Tomte glitches, we find that the joint glitch+signal model is only preferred over the signal-only model in 15\% of our test glitches. However, our study on Tomte glitches also shows that using the Blip trained model improves the signal parameter estimation and reduces the bias in the posteriors. We thus show that although the model selection tests indicate that the Blip model is not a good fit for Tomte glitches, it still improves the signal analysis. This is probably only true due to the similarity in morphology between Blip and Tomte glitches. Further investigations would be needed to draw more conclusions regarding the limits of the glitch model. However, it is straightforward and not computationally expensive to train a new glitch model, and thus, training one model for each glitch class would be the best approach. We thus draw the conclusion that, although a model trained on a different glitch class can be of use, it is better to use a glitch model trained on the same glitch class it will be applied to. If, however, a new type of glitch appears, or there are too few glitches of a certain type to train a new glitch model, making use of a glitch model trained on glitches with similar features would be the next best option.

Another future analysis is to apply the glitch model to real \ac{GW} events contaminated by glitches. This work has focused on proof of concept, and we use injected signals only to demonstrate robustness, since we know their true parameters. The next step would be to analyse real events that are known to be problematic due to glitches and investigate if our model can compare to (and possibly improve) state-of-the-art results. To perform this analysis fully, different glitch models would need to be trained for each class of glitches we want to remove. Some signals of particular interest for glitch subtraction include, for example, GW191109~\cite{Udall:2024ovp} and GW200129~\cite{Macas:2023wiw}.

In conclusion, we have demonstrated the potential of a normalising flow trained glitch model to improve gravitational wave signal analysis and reduce bias through joint inference of signal and glitch. We have also discussed the developments needed to scale our analysis to real events, and future work would include analysis of glitch-contaminated events in the gravitational wave catalogue. 

The code produced and used for this work, including the trained glitch model, is available from~\cite{malz_2025_15316399}.